\documentclass[12pt]{iopart}

\usepackage{hyperref}
\usepackage{acronym}
\usepackage{graphicx}

\graphicspath{{figs/}}

\usepackage{tikz}
\usetikzlibrary{datavisualization}
\usetikzlibrary{datavisualization.formats.functions}

\usepackage{color}
\newcommand{\cboxed}[2][red]{\color{#1}{\color{black}#2}\color{black}}

\newcommand{\TRC}{MOE Key Laboratory of TianQin Mission, TianQin Research Center for Gravitational Physics \& School of Physics and Astronomy, Frontiers Science Center for TianQin, Gravitational Wave Research Center of CNSA, Sun Yat-sen University (Zhuhai Campus), Zhuhai 519082, China}
\newcommand{\bea}{\begin{eqnarray}}
\newcommand{\eea}{\end{eqnarray}}

\newcommand{\bfh}{{\bf h}}

\newcommand{\bfk}{{\bf k}}

\newcommand{\bfp}{{\bf p}}

\newcommand{\bfr}{{\bf r}}

\newcommand{\bfx}{{\bf x}}
\newcommand{\bfy}{{\bf y}}
\newcommand{\bfz}{{\bf z}}
\newcommand{\bfA}{{\bf A}}
\newcommand{\bfB}{{\bf B}}

\newcommand{\bfD}{{\bf D}}

\newcommand{\bfP}{{\bf P}}

\newcommand{\bfR}{{\bf R}}

\newcommand{\htk}{{\hat k}}

\newcommand{\htn}{{\hat n}}

\newcommand{\htp}{{\hat p}}
\newcommand{\htq}{{\hat q}}
\newcommand{\htr}{{\hat r}}

\newcommand{\htu}{{\hat u}}
\newcommand{\htv}{{\hat v}}

\newcommand{\htz}{{\hat z}}

\newcommand{\tdh}{{\td h}}

\newcommand{\tdl}{{\td l}}

\newcommand{\tds}{{\td s}}

\newcommand{\tdy}{{\td y}}

\newcommand{\tdX}{{\td X}}

\newcommand{\mcD}{\mathcal{D}}

\newcommand{\nn}{\nonumber}
\newcommand{\pd}{\partial}
\newcommand{\td}{\tilde}

\def\pdb[#1]#2#3{\left(\frac{\pd #3}{\pd #2}\right)_{#1}}

\long\def\symbolfootnote[#1]#2{\begingroup%
\def\thefootnote{\fnsymbol{footnote}}\footnote[#1]{#2}\endgroup}
\newcommand{\ti}[1]{\tilde{#1}}
\newcommand{\mc}[1]{\mathcal{#1}}
\newcommand{\tbf}[1]{\tilde{\mathbf{#1}}}

\begin{document}

\title{\texttt{GWSpace}: a multi-mission science data simulator for space-based gravitational wave detection}

\author{En-Kun Li$^{1,\dag}$\footnotetext{Co-first author},  
Han Wang$^{1,\dag}$,
Hong-Yu Chen$^1$,
Huimin Fan$^{2,3,4}$,
Ya-Nan Li$^1$,
Zhi-Yuan Li$^1$,
Zheng-Cheng Liang$^5$,
Xiang-Yu Lyu$^1$,
Tian-Xiao Wang$^1$,
Zheng Wu$^6$,
Chang-Qing Ye$^1$,
Xue-Ting Zhang$^1$,
Yiming Hu$^1$,
Jianwei Mei$^1$}
\ead{meijw@sysu.edu.cn (J. Mei, Corresponding Author)}
\address{$^1$\TRC}
\address{$^2$Department of Physics, Hebei University, Baoding, 071002, China\\
$^3$Hebei Key Laboratory of High-precision Computation and Application of Quantum Field Theory, Baoding, 071002, China \\
$^4$Hebei Research Center of the Basic Discipline for Computational Physics, Baoding, 071002, China}
\address{$^5$School of Physics, Henan Normal University, Xinxiang 453007, China}
\address{$^6$Laboratoire des 2 Infinis - Toulouse (L2IT-IN2P3), Universit\'{e} de Toulouse, CNRS, F-31062 Toulouse Cedex 9, France  }


\begin{abstract}
Space-based gravitational wave detectors such as TianQin, LISA, and Taiji have the potential to outperform themselves through joint observation.
To achieve this, it is desirable to practice joint data analysis in advance on simulated data that encodes the intrinsic correlation among the signals found in different detectors that operate simultaneously.
In this paper, we introduce \texttt{\href{https://github.com/TianQinSYSU/GWSpace}{GWSpace}}, a package that can simulate the joint detection data from TianQin, LISA, and Taiji.
The software is not a groundbreaking work that starts from scratch.
Rather, we use as many open-source resources as possible, tailoring them to the needs of simulating the multi-mission science data and putting everything into a ready-to-go and easy-to-use package.
We shall describe the main components, the construction, and a few examples of application of the package.
A common coordinate system, namely the Solar System Barycenter (SSB) coordinate system, is utilized to calculate spacecraft orbits for all three missions.
The paper also provides a brief derivation of the detection process and outlines the general waveform of sources detectable by these detectors.
\end{abstract}


\maketitle
\acrodef{SSB}{solar system barycenter}
\acrodef{ssb}[SSB]{solar system barycenter}
\acrodef{ICRS}{International Celestial Reference System}
\acrodef{GR}{general relativity}
\acrodef{GW}{gravitational wave}
\acrodef{gw}[GW]{gravitational wave}
\acrodef{CMB}{cosmic microwave background}
\acrodef{bbh}[BBH]{Binary Black Hole}
\acrodef{bhb}[BHB]{black hole binariy}
\acrodef{gcb}[GCB]{Galaxy Compact Binary}
\acrodef{mbhb}[MBHB]{Massive Black Hole Binary}
\acrodef{sbhb}[SBHB]{Stellar-mass Black Hole Binary}
\acrodef{emri}[EMRI]{Extreme Mass Ratio Inspirals}
\acrodef{sgwb}[SGWB]{Stochastic Gravitational Wave Background}
\acrodef{tdi}[TDI]{Time Delay Interferometry}
\acrodef{lisa}[LISA]{Laser Interferometer Space Antenna}
\acrodef{mhz}[mHz]{milli-Hertz}
\acrodef{gbd}[GBD]{ground-based GW detector}
\acrodef{sbd}[SBD]{space-based GW detector}
\acrodef{psd}[PSD]{power spectra density}
\acrodef{SPA}{stationary phase approximation}
\acrodef{au}[AU]{astronomical unit}
\acrodef{wdb}[WDBs]{white dwarf binaries}
\tableofcontents

\section{Introduction}

Several space-based \ac{gw} detectors, including TianQin \cite{Luo:2015ght}, the \ac{lisa} \cite{Danzmann:1997hm, LISA:2017pwj}, and Taiji \cite{Gong:2014mca, Hu:2017mde}, are eyeing for launch around mid-2030s.
These detectors will, for the first time, open the unexplored \ac{mhz} frequency band of \ac{gw} spectrum.
In complement to the current \acp{gbd} \cite{LIGOScientific:2021djp},
\acp{sbd} enjoy a plethora of new types of sources, including the \acf{gcb} \cite{Huang:2020rjf, Lu:2022ywf},
the \ac{mbhb} \cite{Wang:2019ryf},
the \ac{sbhb} \cite{Liu:2020eko, Wang2023},
the \ac{emri} \cite{Fan:2020zhy, Zhang:2022xuq},
the \ac{sgwb} \cite{Liang:2021bde, Cheng:2022vct},
and etc \cite{LISA:2022yao, LISA:2022kgy, LISACosmologyWorkingGroup:2022jok, LISACosmologyWorkingGroup:2022wjo, LISACosmologyWorkingGroup:2022kbp, LISACosmologyWorkingGroup:2019mwx}.

Unlike \acp{gbd} that mainly capture \ac{gw} events in their short-lived merger phases, \acp{sbd} detect \ac{gw} events mostly during their long-lasting inspiral phases, resulting in complex data sets with overlapping signals in time and frequency.
Consequently, this poses significant challenges to data analysis \cite{MockLISADataChallengeTaskForce:2009wir, Baghi:2022ucj, Ren:2023yec,Wang2023}.
So several mock data challenges, such as mock LISA data challenge (MLDC) \cite{arnaud2007report, MockLISADataChallengeTaskForce:2007iof, MockLISADataChallengeTaskForce:2009wir}, which is now replaced by LISA data challenge (LDC) \cite{Baghi:2022ucj}, and Taiji data challenge (TDC) \cite{Ren:2023yec}, have been set up to help develop the necessary tools need for space-based \ac{gw} data analysis.

It is possible that more than one of the detectors, TianQin, LISA, and Taiji, will be observing concurrently during the mid-2030s, enabling a network approach to detect some \ac{gw} signals.
These detectors can then observe the same \ac{gw} signals from different locations in the solar system, effectively forming a virtual detector with a much larger size \cite{Gong:2021gvw}, leading to significant improvements in sky localization accuracy \cite{Crowder:2005nr, Ruan:2020smc, Wang:2020vkg, Zhu:2021bpp, Lyu:2023ctt}, allowing for the discovery of more sources and a deeper understanding of physics \cite{Liang:2022ufy, Liang:2023fdf}.
More examples showing how joint detection can improve over individual detectors can be found in \cite{Ruan:2020smc, Schutz:2011tw, Lyu:2023ctt, Huang:2020rjf, Wang:2019ryf, Liu:2020eko, Fan:2020zhy}. A comprehensive study of how the joint detection with TianQin and LISA can improve over each detector can be found in \cite{Torres-Orjuela:2023hfd}.
What's more, the difficulties faced by space-based \ac{gw} data analysis are partially due to parameter degeneracy \cite{Xie:2022brn}, and it has been shown that joint detection can also be helpful here by breaking some of the degeneracies \cite{Roulet:2022kot}. So it is important to seriously consider the possibility of doing joint data analysis from different \ac{sbd} combinations.

There are challenges in conducting data analysis for joint observations with multiple detectors. For example, due to the significant differences in arm lengths and orbits, approximations and optimized algorithms developed for geocentric and heliocentric cannot be directly applied interchangeably. What's more, variations in the separations among the detectors will affect the correlation of the signals, and this requires comprehensive consideration in the calculation of the likelihood and covariance matrices. To facilitate the study of problems involved in the analysis of joint observational data, we introduce in this paper \texttt{\href{https://github.com/TianQinSYSU/GWSpace}{GWSpace}}, which is a package that can simulate the joint simulation data from all three \acp{sbd} mentioned above.

Although MLDC, LDC, and TDC have already achieved simulating data for individual detectors like LISA and Taiji, there are new problems to be solved when one wants to simulate data for all three detectors operating together.
For example, due to the shorter arm-length of TianQin, its sensitivity frequency band is more shifted toward the higher frequency end, ranging from $10^{-4}$ to $1$ Hz \cite{Luo:2015ght}, as compared to about $[2\times10^{-5}, 10^{-1}]$ Hz for \ac{lisa} and Taiji \cite{LISA:2017pwj, Hu:2017mde}.
Because of this, the response model derived using the low-frequency limit method \cite{Cutler:1997ta} that works for LISA and Taiji is not always valid for TianQin.
Therefore, it is necessary to consider the full-frequency response models to accurately describe the behavior of all the detectors across the entire frequency spectrum \cite{Marsat:2018oam, Marsat:2020rtl}.
Another issue is that one needs to study the response of the three \acp{sbd} by using the same coordinate system to correctly reveal the correlation among them.
The \ac{ssb} coordinate system is identified as the most straightforward choice for this purpose.

The paper is structured as follows.
Section \ref{sec:coord} specifies the coordinate systems used in this paper.
Section \ref{sec:detector} specifies the orbits of the three \acp{sbd} involved, namely TianQin, LISA and Taiji.
Section \ref{sec:response}, \ref{sec:TDI}, and \ref{sec:source} detail the response, \ac{tdi} combinations, and source waveforms used in \texttt{GWSpace}. 
Some example data sets are described in section \ref{sec:data_set}.
A summary is in section \ref{sec:sum}.

\section{Coordinate systems}
\label{sec:coord}



Two basic coordinate systems will be used in this paper: the astronomical ecliptic coordinate system used to describe the detector, hence called the detector frame, and the coordinate system adapted to the description of \ac{GW} sources, hence called the source frame.

\begin{figure}[h]
    \centering
    \includegraphics[width=0.8\linewidth]{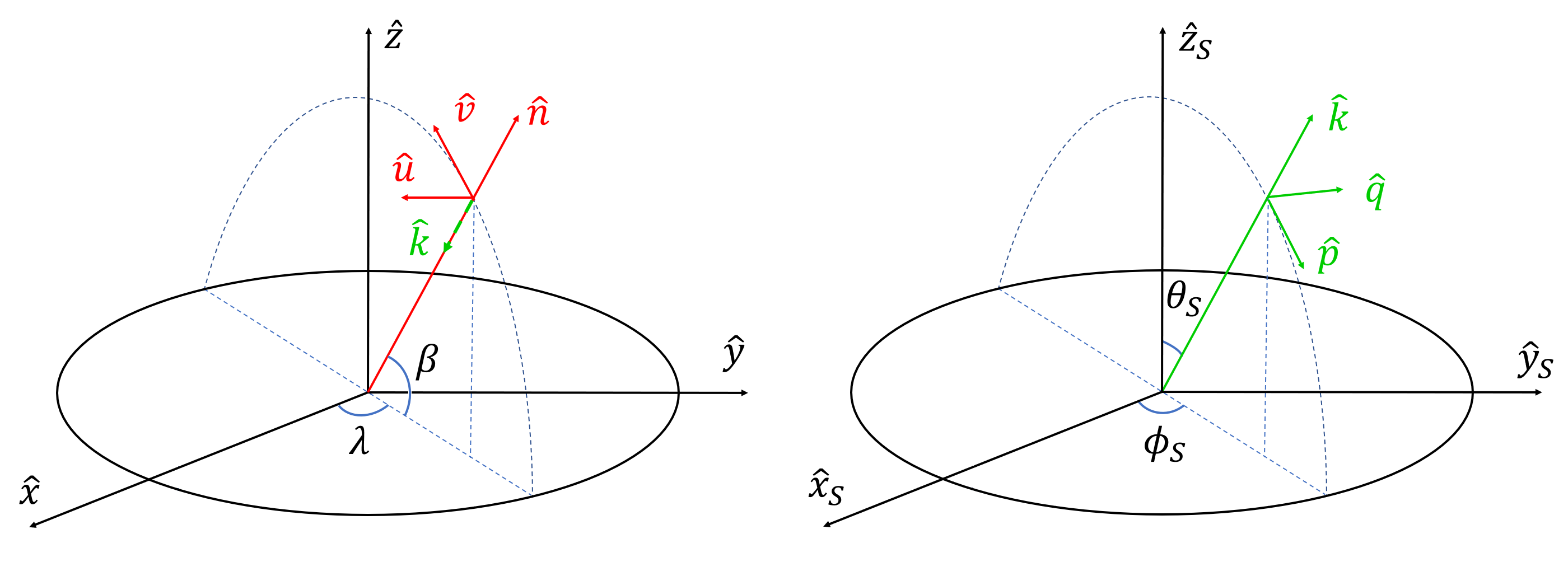}
    \caption{(Left) The detector frame and the relative orientation between $(\hat{x}, \hat{y}, \hat{z})$ and $(\hat{ n}, \hat{u}, \hat{v})$. (Right) The source frame and the relative orientation between $(\hat{x}_S, \hat{y}_S, \hat{z}_S)$ and $(\htk, \hat{p}, \hat{ q})$.}
    \label{fig:coors}
\end{figure}

The detector frame, as illustrated in Fig. \ref{fig:coors} (Left), is defined with the origin at the \ac{SSB}.
In this frame, the $z$-axis is oriented perpendicular to the ecliptic and points towards the north,
while the $x$-axis points towards the March equinox.
The $y$-axis is obtained as $\bfy= \bfz \times \bfx\,$.
The direction to a \ac{GW} source is indicated with the unit vector $\htn=\htn(\lambda,\beta)$, 
where $\lambda$ and $\beta$ are the celestial longitudes and celestial latitude of the source, respectively. 
To describe the polarization of \acp{GW} propagating along $\htk=-\htn\,$, two additional auxiliary unit vectors are introduced
\footnote{\url{https://lisa-ldc.lal.in2p3.fr/static/data/pdf/LDC-manual-002.pdf}}
\begin{equation}
    \hat{u} = \frac{\htn\times\hat{z}}{|\htn\times\hat{z}|}
    = \frac{\htz \times \htk}{| \htz \times \htk |}
    \,,\quad
    \hat{v}=\hat{u}\times\htn
    = \htk \times \htu \,,
\end{equation}
so that the trio, $(\hat{u}, \hat{v}, \htk)\,$, forms a right-handed orthogonal basis.
Then, one can obtain that
\begin{eqnarray}
    \htu =& \left[ \sin{\lambda}, \  - \cos\lambda, \  0\right], \\
    \htv =& \left[ - \sin\beta \cos\lambda, \  - \sin\beta \sin\lambda, \  \cos\beta\right], \\
    \htk =& = -\htn = \left[ - \cos\beta \cos\lambda, \ - \sin\lambda \cos\beta, \ - \sin\beta\right].
\end{eqnarray}

The source frame is illustrated in Fig. \ref{fig:coors} (Right). Exactly how the origin and the axes, $(\hat{x}_S, \hat{y}_S, \hat{z}_S)$, are chosen for the source dynamics will be determined on a case-by-case basis. The direction to the \ac{GW} detector is indicated with the unit vector $\htk=\htk(\theta_S,\phi_S)\,$. To describe the polarization of \acp{GW} propagating along $\htk\,$, two auxiliary unit vectors are also introduced,
\bea\hat{q}=\frac{\hat{z}\times\htk}{|\hat{z}\times\htk|}\,,\quad\hat{p}=\hat{q}\times\htk\,,\eea
so that the trio, $(\hat{p}, \hat{q}, \htk)$, forms a right-handed orthogonal basis.
Then, one can obtain that
\begin{eqnarray}
    \htp =& \left[ \cos\iota \cos\varphi, \  \sin\varphi \cos\iota, \  - \sin\iota\right], \\
    \htq =& \left[ - \sin\varphi, \  \cos\varphi, \  0\right], \\
    \htk = & \left[ \sin\iota \cos\varphi, \  \sin\iota \sin\varphi, \  \cos\iota\right].
\end{eqnarray}

\begin{figure}[h]
    \centering
    \includegraphics[width=0.3\linewidth]{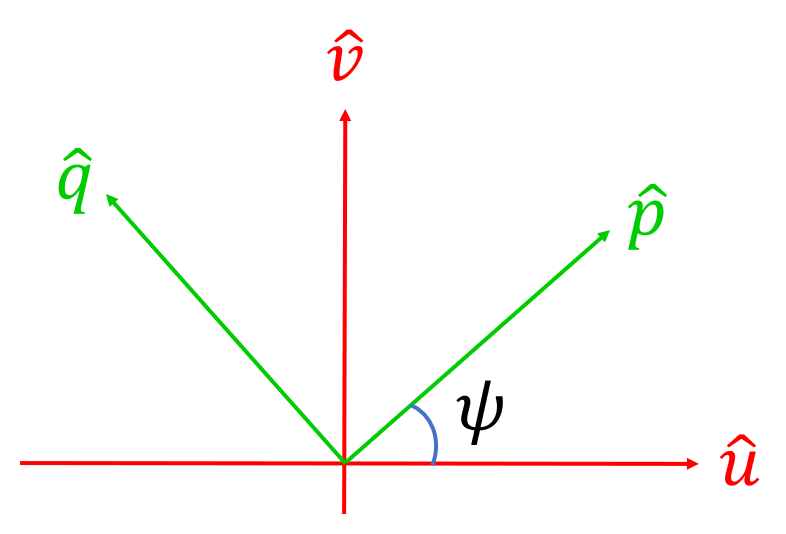}
    \caption{Relative orientations of $(\hat{u}, \hat{v})$ and $(\hat{p}, \hat{q})\,$. Note that $\htk$ is perpendicular to and coming out of the paper.}
    \label{fig:coors3}
\end{figure}

Since $\htn=-\htk\,$, the planes spanned by $(\hat{u}, \hat{v})$ and $(\hat{p}, \hat{q})$ are parallel to each other (see Fig. \ref{fig:coors3}). As a result,
\begin{eqnarray}
    \hat{p} =& \cos\psi\, \hat{u} + \sin\psi\, \hat{v}\,,
    \quad
    \hat{q}=- \sin\psi\, \hat{u} + \cos\psi\, \hat{v}\,,
    \nn \\
    \hat{u} =& \cos\psi\, \hat{p} - \sin\psi\, \hat{q}\,,
    \quad
    \hat{v}= \sin\psi\, \hat{p} + \cos\psi\, \hat{q}\,.
\end{eqnarray}
Thus, the polarization angle can be computed as
\begin{equation}
    \psi = \mathrm{arctan}_2 \left[ \htp \cdot \htu, \htp \cdot \htv \right].
\end{equation}

Using the basis vectors, one can define the polarization tensors in the source frame and the SSB frame.
In the source frame
\begin{equation}
    e^+_{ij} = (\htp \otimes \htp - \htq \otimes \htq)_{ij},
    \quad
    e^\times_{ij} = (\htp \otimes \htq + \htq \otimes \htp)_{ij}.
\end{equation}
Similarly, in the SSB frame
\begin{equation}
    \epsilon^+_{ij} = (\htu \otimes \htu - \htv \otimes \htv)_{ij},
    \quad
    \epsilon^\times_{ij} = (\htu \otimes \htv + \htv \otimes \htu)_{ij}.
\end{equation}
After some calculation, the relation between the polarization tensors can be rewritten as
\begin{eqnarray}
    e^+ 
    &=& \epsilon^+ \cos2\psi + \epsilon^\times \sin 2\psi,
    \\
    e^\times 
    &=& - \epsilon^+ \sin 2\psi + \epsilon^\times \cos 2\psi.
\end{eqnarray}
In the source frame, the GW strain in a transverse-traceless gauge takes the form
\begin{equation}
    h_{ij}^{TT} = e^+_{ij} h_+ + e^\times_{ij} h_\times,
\end{equation}
where $h_{+,\times}$ are the plus and cross mode of \ac{gw}.
The corresponding representation of the strain in the SSB frame is
\begin{eqnarray}
    h_{ij}^{TT} &=\, & h_+ (\epsilon^+ \cos 2\psi + \epsilon^\times \sin 2\psi)_{ij}
        + h_\times (\epsilon^\times \cos 2\psi - \epsilon^+ \sin 2\psi)_{ij} \nonumber \\
        &=\, & (h_+ \cos 2\psi - h_\times \sin 2\psi) \epsilon^+_{ij}
        + (h_+ \sin 2\psi + h_\times \cos 2\psi) \epsilon^\times_{ij}.
\end{eqnarray}
Then, in the SSB frame, one can define two new modes of \ac{gw} as
\begin{eqnarray}
        h^{\rm SSB}_+ =\, & h_+ \cos 2\psi - h_\times \sin 2\psi, \\
        h^{\rm SSB}_\times =\, & h_+ \sin 2\psi + h_\times \cos 2\psi.
\end{eqnarray}

\section{Detectors}
\label{sec:detector}

\begin{figure}[htbp]
    \centering
    \includegraphics[width=0.6\linewidth]{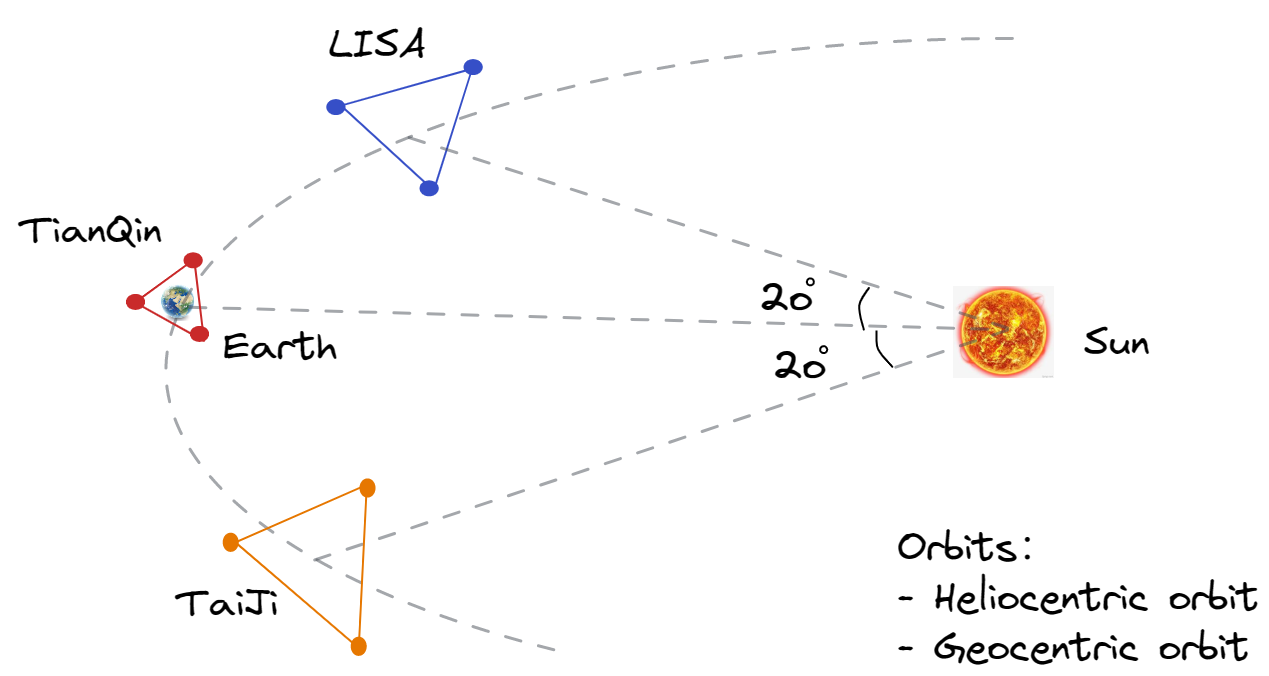}
    \caption{Schematic of the spacecraft's orbit in the \ac{ssb} coordinate system.}
    \label{fig:SBD_Orbit_SSB}
\end{figure}

The three \acp{sbd}, TianQin, \ac{lisa}, and Taiji, all consist of three identical spacecraft that form a nearly equilateral triangle.
The main difference is in their orbits: TianQin is placed on nearly identical nearly circular geocentric orbits with a radius of about $10^5$ km \cite{Luo:2015ght}.
The detector plane of TianQin is directed towards the calibration source \texttt{RX J0806.3+1527}.
In contrast, \ac{lisa} and Taiji are placed on Earth-like heliocentric orbits with a semi-major axis of about 1 \ac{au} from the Sun \cite{LISA:2017pwj, Hu:2017mde}, and their detector plane rotates around in a yearly cycle.
The arm length of TianQin is about $1.7\times 10^5$ km, while those of \ac{lisa} and Taiji are about $2.5\times 10^6$ km and $3\times 10^6$ km, respectively.
The center of \ac{lisa} is approximately $20$ degrees behind the Earth, while that of Taiji is approximately $20$ degrees ahead of the Earth \cite{Gong:2021gvw} (see Fig.~\ref{fig:SBD_Orbit_SSB}).
By selecting a geocentric orbit, TianQin can transmit data back to Earth in nearly real-time, making it more adapted to multi-messenger astronomy \cite{Chen:2023qga}.

In the following subsections, we utilize the Keplerian orbit to approximate the motion of the spacecraft in the \ac{ssb}.

\subsection{TianQin: geocentric orbit}

\begin{figure}[htbp]
    \centering
    \includegraphics[width=0.8\linewidth]{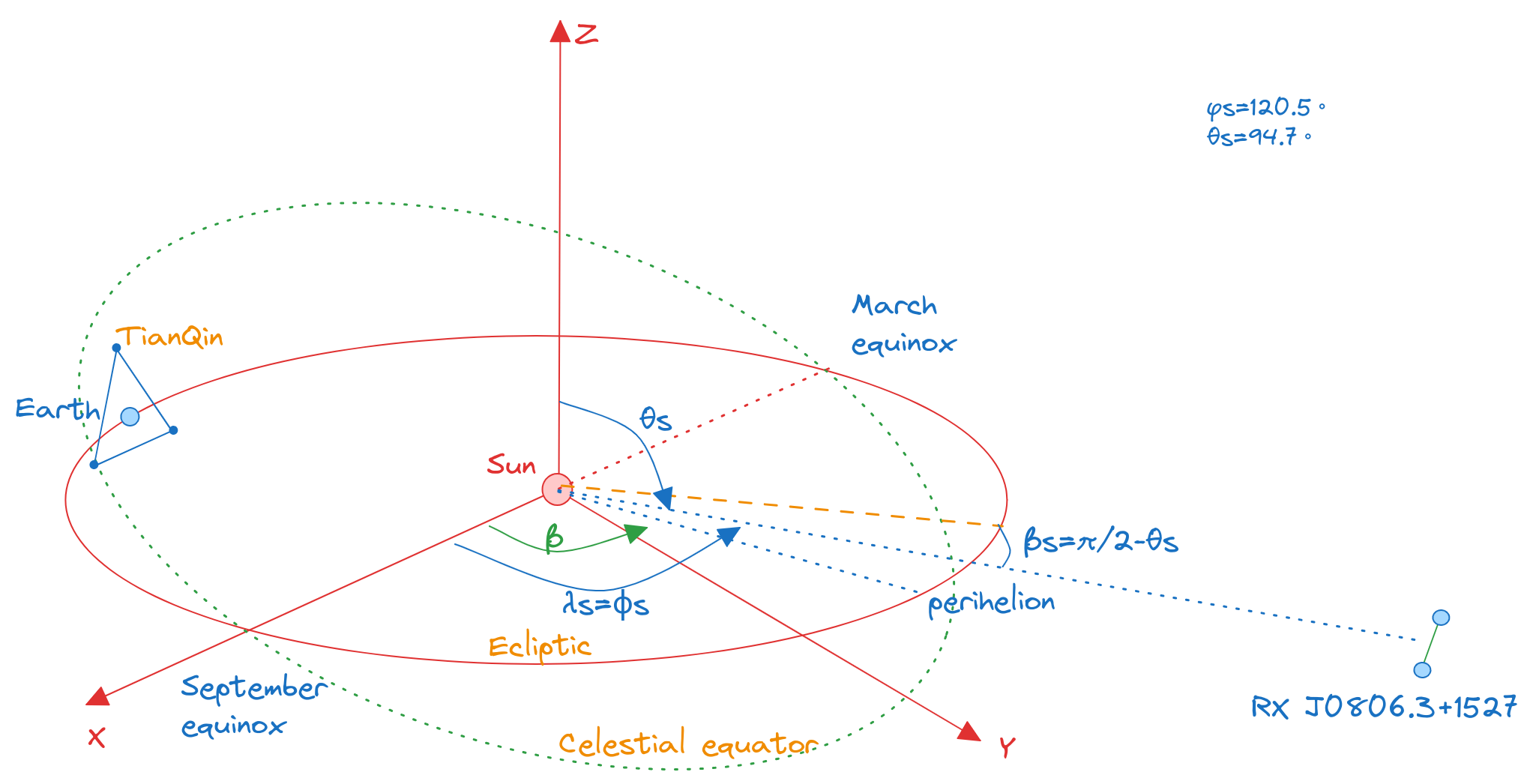}
    \caption{Schematic of the TianQin spacecraft's orbit for TianQin in the 
    \ac{SSB} coordinate system.}
    \label{fig:TianQin_Orbit_SSB}
\end{figure}

In Fig.~\ref{fig:TianQin_Orbit_SSB}, we present a schematic of the spacecraft orbits for TianQin.
The $x$-axis is defined as the direction from the Sun to the September equinox, while the $z$-axis represents the angular momentum direction of the Earth.
For detailed information on the derivatives for the Keplerian orbit of TianQin, please refer to Ref.~\cite{Hu:2018yqb}.

The following presents a simplified and non-realistic depiction of the orbit, focusing on the motion of the Earth's center or guiding center in the \ac{ssb} frame:
\begin{eqnarray}
    X(t) 
    &= & R \bigg[ \cos(\alpha - \beta) - e (1 + \sin^2 (\alpha - \beta))
    \nonumber\\
    && - \frac{3}{2} e^2 \cos(\alpha - \beta) \sin^2(\alpha - \beta) \bigg],
    \label{eq:earth_x}
    \\
    Y(t) 
    &= & R \bigg[ \sin(\alpha - \beta) + e \sin(\alpha -\beta) \cos(\alpha -\beta)
    \nonumber\\
    && + \frac{1}{2} e^2 \sin(\alpha - \beta) (1-3\sin^2(\alpha - \beta)) \bigg],
    \label{eq:earth_y}
    \\
    Z(t) &= & 0,
    \label{eq:earth_z}
\end{eqnarray}
where $\alpha = 2\pi f_m t + \kappa_0$,
$f_m = 1/(\rm one~ sidereal~ year) = 3.14 \times 10^{-8}$ Hz is the orbit modulation frequency,
$\kappa_0$ is the mean ecliptic longitude measured from the vernal equinox (or September equinox) at $t=0$,
and $\beta$ denotes the angle measured from the vernal equinox to the perihelion.

\begin{figure}
    \centering
    \includegraphics[width=0.8\linewidth]{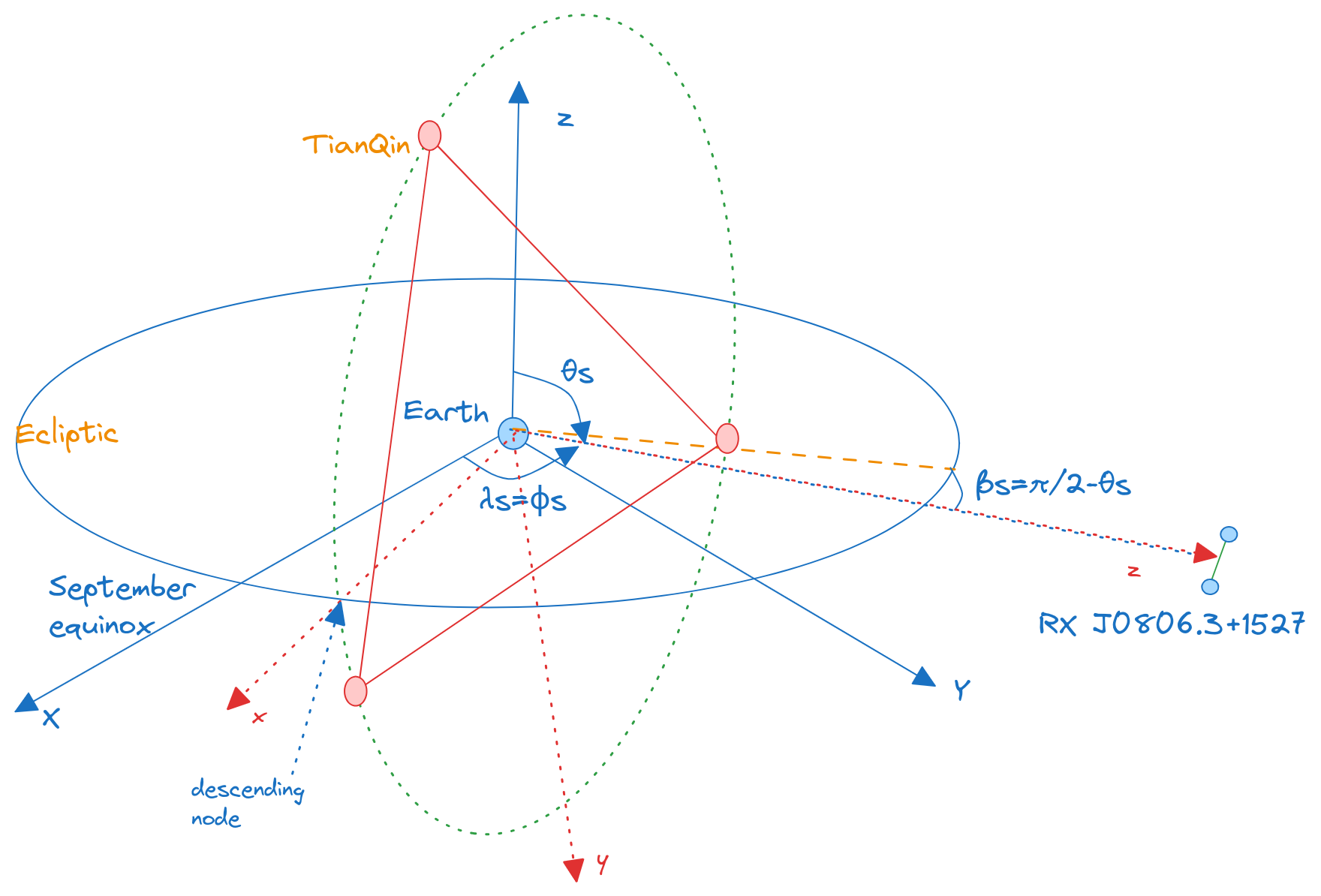}
    \caption{Schematic of the detector coordinate system $\{\ti{x}, \ti{y}, \ti{z}\}$ and the geocentric-ecliptic coordinate system $\{x,y,z\}$.
    $\ti{x}$ point to the descending node, $\ti{z}$ axis points to J0806.}
    \label{fig:TianQin_Orbit_Local}
\end{figure}

In the context of the TianQin spacecraft's motion around the Earth, its orbits remain consistent with Eq.~(\ref{eq:earth_z}).
However, when considering the \ac{ssb} frame, TianQin assumes a specific orientation towards the direction of J0806 ($\{\lambda_s, \beta_s \} = \{120.5^\circ, -4.7^\circ\}$, as shown in Fig.~\ref{fig:TianQin_Orbit_Local}).
Introducing a coordinate system rotation and disregarding eccentricity, the description of TianQin's orbits can be further refined \cite{Hu:2018yqb}
\begin{eqnarray}
    x_n =& \frac{L}{\sqrt{3}} \left[ \sin\beta_s \cos\lambda_s \sin(\alpha_n - \beta') + \sin\lambda_s \cos(\alpha_n - \beta')\right], \\
    y_n =& \frac{L}{\sqrt{3}} \left[ \sin\beta_s \sin\lambda_s \sin(\alpha_n - \beta') - \cos\lambda_s \cos(\alpha_n - \beta')\right], \\
    z_n =& -\frac{L}{\sqrt{3}} \cos\beta_s \sin(\alpha_n -\beta'),
\end{eqnarray}
where $L = \sqrt{3} \, R_{tq}$ is the arm-length between the two spacecraft,
$\alpha(t) = 2\pi f_{sc} t + \kappa_n$,
$\kappa_n = + \frac{2}{3} \pi (n-1) + \lambda$,
$\lambda$ is the initial orbit phase of the first ($n=1$) spacecraft measured from $\ti{x}$ axis,
$f_{sc} = 1/\sqrt{G M_{Earth}/R_{tq}^3} \simeq 1/(3.65 \,\rm day)$ is the modulation frequency due to the rotation of the detector around the guiding center,
$\beta'$ is the angle measured from the $\ti{x}$ axis to the perigee of the first spacecraft orbit.
Here, we assume the three spacecraft are in circular orbits around the Earth, thus, $\beta'$ will be some arbitrary number (one can just set it as $0$).

\subsection{LISA and Taiji: heliocentric orbit}



According to the description in Ref.~\cite{Rubbo:2003ap}, when considering a constellation of spacecraft in individual Keplerian orbits with an inclination of $\iota = \sqrt{e}$,  the coordinates of each spacecraft can be elegantly expressed in the following form (this expression has been expanded up to the second order of eccentricity) \cite{Rubbo:2003ap}
\begin{eqnarray}
    x_n =& a \cos(\alpha'')
    + a e \left( \sin\alpha'' \cos\alpha'' \sin\beta_n' - (1+\sin^2\alpha'') \cos\beta_n' \right)
    \nonumber \\
    & + \frac{1}{8} a  e^2 \left( 3\cos(3\alpha'' - 2\beta_n') - 10 \cos\beta_n' - 5 \cos(\alpha'' - 2\beta_n') \right),
    \label{eq:lisa_x}
    \\
    y_n =& a \sin(\alpha'') + a e \left( \sin\alpha'' \cos\alpha'' \cos \beta_n' - (1+\cos^2\alpha'') \sin\beta_n' \right)
    \nonumber \\
    & + \frac{1}{8} a e^2 \left( 3\sin (3\alpha'' -2\beta_n') -10 \sin\alpha'' + 5\sin(\alpha''-2\beta_n') \right),
    \label{eq:lisa_y}
    \\
    z_n =& -\sqrt{3} a e \cos(\alpha'' -\beta_n')
    +\sqrt{3} ae^2 \left[ 1 + \sin^2 (\alpha'' -\beta_n') \right].
    \label{eq:lisa_z}
\end{eqnarray}
Here $a = R_{LISA, TJ} = 1$ \ac{au} is the radial distance to the guiding center for \ac{lisa} and Taiji,
$\alpha'' = \alpha - \beta \mp 20^\circ$ for \ac{lisa} and Taiji,
where $\alpha$ and $\beta$ are the same as those in Earth orbit or Eqs.~(\ref{eq:earth_x})-(\ref{eq:earth_z}).
And $\beta_n' = \frac{2\pi}{3} (n-1) + \lambda'$,
$\lambda'$ is the initial orientation of the constellation,
$e \simeq L_{LISA, TJ}/(2 a \sqrt{3})$ represent the orbital eccentricity,
$L_{LISA} = 2.5\times 10^6$ km and $L_{TJ} = 3\times 10^6$ km is the arm-length between two spacecraft for \ac{lisa} and Taiji, respectively.

\begin{figure}[htbp]
    \centering
    \includegraphics[width=0.65\linewidth]{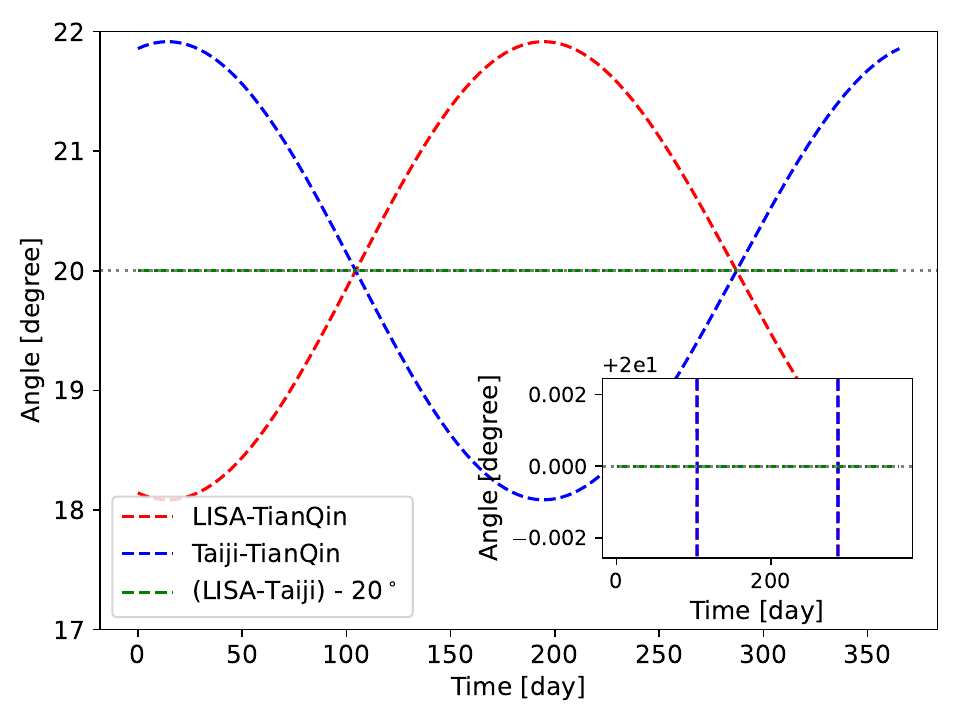}
    \caption{The relative angle between different detectors.
    Here, for improved visual clarity, the angle between \ac{lisa} and Taiji has been adjusted by subtracting 20 degrees.}
    \label{fig:angle_sdb}
\end{figure}

While the spacecraft orbits for \ac{lisa} and Taiji are situated in the ecliptic planes, their constellations' guiding center follows nearly circular trajectories.
In the \texttt{GWSpace} code, the perihelion angle of the three spacecraft for \ac{lisa} and Taiji is set to be the same as that of Earth.
However, TianQin's guiding center coincides with Earth's, resulting in the changing angle between \ac{lisa} and Taiji over time.
Figure~\ref{fig:angle_sdb} illustrates the relative angles between the different detectors.
It can be observed that the angle between \ac{lisa} or Taiji and Earth varies between $18^\circ$ and $22^\circ$, while the angle between \ac{lisa} and Taiji is approximately $40^\circ$, with a slight variation of around $2.4 \times 10^{-3}$.
These findings are consistent with the proposed orbit described in Ref.~\cite{LISA:2017pwj, Hu:2017mde}.

\section{Detector response}
\label{sec:response}

In a vacuum, propagating \acp{gw} induces a time-varying strain in the fabric of space-time.
This strain can alter the proper distance between freely falling masses, providing a means to gather information about the \acp{gw}.
One approach is to measure the variation in light travel time or optical path length between two test masses \cite{Jolien:GW_phy_astr2011}.
As a \ac{gw} passes through, these separated masses will experience relative acceleration or tilting.
Consequently, a \ac{gw} detector is employed to monitor the separation between the test masses.
There are two commonly used methods to monitor the distance between two objects: radar ranging or similar techniques, and measuring the Doppler shift in a signal transmitted from one object to the other \cite{Jolien:GW_phy_astr2011}.
However, a question arises regarding whether the \ac{gw} affects the electromagnetic waves used for measuring distances \cite{Jolien:GW_phy_astr2011}.
In the following sections, we will provide a brief overview of how a \ac{gw} detector responds to \ac{gw} signals.

\subsection{The general waveform and mode decomposition}
\label{sec:basic_def}

Assuming a universe consisting solely of vacuum and \ac{gw}.
Since \acp{gw} are very weak, the metric of the spacetime perturbed by a \ac{gw} can be described as
\begin{eqnarray}
    ds^2 &=& -c^2 dt^2 + \left[ \delta_{ij} +h_{ij} (t) \right] dx^i dx^j,
\end{eqnarray}
where $h_{ij}$ is the tensor perturbation, it is directly related to the \ac{gw} itself, carrying information about its amplitude, frequency, and polarization.
By analyzing the changes in the metric caused by the \ac{gw}, we can extract valuable information about the \ac{gw} signal.
In the TT coordinate system (with coordinates $x^0 = ct, x^1 = x, x^2 = y, x^3 = z$), a weak \ac{gw} can be described as a weak plane wave traveling in the $+z$ direction.
The line element describing the metric of spacetime in this scenario is given by
\begin{eqnarray}
    ds^2  &=& -c^2 dt^2
    + \left( 1 + h_+ \left( t - \frac{z}{c} \right) \right) dx^2
    + \left( 1 - h_+ \left( t - \frac{z}{c}\right) \right) dy^2
    \nonumber \\
    &&+ 2h_\times \left( t- \frac{z}{c} \right)dx dy
    + dz^2.
\end{eqnarray}

\paragraph{General waveform}
The \ac{gw} can be approximated as an arbitrary plane wave with wave vector $\vec{k}$ and a Tensor form `amplitude', thus
\begin{equation}
    \bfh(t, \bfr) = \bfh_0 e^{\rmi (2\pi f t - \vec{k} \cdot \bfr/c)}
    = \bfh_0 e^{\rmi 2\pi f (t- \htk \cdot \bfr/c)}
    = h(t- \htk \cdot \bfr ) = h(\xi),
\end{equation}
where $\htk = \frac{\vec{k}}{|\vec{k}|} = \frac{\vec{k}}{2\pi f}$ is the propagation direction of \ac{gw},
$\bfr$ is an arbitrary direction,
$\xi = t - \htk \cdot \bfr$ is a surface of constant phase.

There is relative motion between the source frame and the detector frame.
In the detector frame, the \ac{SSB} is moving relative to the \ac{CMB} with a peculiar velocity $v\approx 370$ km/s $\approx 0.0012$, along the direction $\lambda\approx172^\circ\,$, $\beta\approx-11^\circ$ \cite{Kogut:1993ag}.
In the source frame, the velocities of the sources can be introduced as model parameters.

\paragraph{Mode decomposition}
In the source frame, the gravitational wave can be further decomposed using spin-weighted spherical harmonics \cite{Goldberg:1966uu} ${}_{-2} Y_{\ell m} (\iota, \varphi)$ as
\begin{equation}
    h_+ - \rmi h_\times = \sum_{\ell \geq 2} \sum_{m = -\ell}^{\ell} {}_{-2} Y_{\ell m} (\iota, \varphi) h_{\ell m}.
\end{equation}
where $\{\iota, \varphi\}$ represent the inclination and phase describing the orientation of emission.
The primary harmonic is $h_{22}$, while the others are called higher harmonics or higher modes.
And each mode can be described as
\begin{equation}
    h_{\ell m} = A_{\ell m} e^{-\rmi \Phi_{\ell m}}.
\end{equation}
Based on this decomposition, we obtain
\begin{eqnarray}
    h_+ = & \frac{1}{2} \sum_{\ell, m} \left( {}_{-2} Y_{\ell m} (\iota, \varphi) h_{\ell m} + {}_{-2} Y_{\ell m}^* (\iota, \varphi) h_{\ell m}^* \right),
    \\
    h_\times = & \frac{\rmi}{2} \sum_{\ell, m} \left( {}_{-2} Y_{\ell m} (\iota, \varphi) h_{\ell m} - {}_{-2} Y_{\ell m}^* (\iota, \varphi) h_{\ell m}^* \right).
\end{eqnarray}

In particular, for the non-processing binary systems with a fixed equatorial plane of orbit, there exists an exact symmetry relation between modes
\begin{equation}
    h_{\ell, -m} = (-1)^{\ell} h^*_{\ell m}.
    \label{eq:h_lm}
\end{equation}
With this symmetry, one has
\begin{equation}
    h_{+, \times} = \sum_{\ell, m} K_{\ell m}^{+,\times} h_{\ell m},
\end{equation}
where
\begin{eqnarray}
    K_{\ell m}^+ &=& \frac{1}{2} \left( {}_{-2} Y_{\ell m} + (-1)^\ell {}_{-2} Y_{\ell, -m}^* \right),
    \\
    K_{\ell m}^\times &=& \frac{\rmi}{2} \left( {}_{-2} Y_{\ell m} - (-1)^\ell {}_{-2} Y_{\ell, -m}^* \right).
\end{eqnarray}


It is convenient to introduce mode-by-mode polarization matrices
\begin{equation}
    P_{\ell m} = e_+ K_{\ell m}^+ + e_\times K_{\ell m}^\times,
\end{equation}
so that the \ac{gw} signal in matrix form will be
\begin{equation}
    \bfh^{TT} = \sum_{\ell, m} P_{\ell m} h_{\ell m}.
\end{equation}
In the \ac{ssb} frame, one can write
\begin{equation}\label{eq:pppc}
    P_+ + \rmi P_\times = e^{-\rmi 2\psi} (\epsilon_+ + \rmi \epsilon_\times).
\end{equation}
With the above equations, $P_{\ell m}$ will be
\begin{eqnarray}\label{eq:plm}
    P_{\ell m} (\iota, \varphi, \psi) &=&
    \frac{1}{2} {}_{-2} Y_{\ell m} (\iota, \varphi)
    e^{- \rmi 2 \psi} (\epsilon_+ + \rmi \epsilon_\times)
    \nonumber \\
    && + \frac{1}{2} (-1)^{\ell} {}_{-2} Y_{\ell, -m}^* (\ell, \varphi)
    e^{+\rmi 2\psi} (\epsilon_+ - \rmi \epsilon_\times).
\end{eqnarray}
In this way, we can factor out explicitly all dependencies in the extrinsic parameters $(\iota, \varphi, \psi)$.

Suppose that the \ac{gw} only has the main mode, i.e., the 22 mode, we have $h_{22} = A_{22} e^{-\rmi \Phi_{22}}$ and $h_{2,-2} = h_{22}^* = A_{22} e^{\rmi \Phi}$.
The expressions of the spin-weighted spherical harmonics for the mode of $\{2,\pm 2\}$ are
\begin{equation}
    {}_{-2} Y_{22}(\iota, \varphi) =
    \frac{1}{2} \sqrt{\frac{5}{\pi}} \cos^4 \frac{\iota}{2} e^{\rmi 2 \varphi},
    \qquad
    {}_{-2} Y_{2,-2}(\iota, \varphi) =
    \frac{1}{2} \sqrt{\frac{5}{\pi}} \sin^4 \frac{\iota}{2} e^{-\rmi 2 \varphi},
\end{equation}
and
\begin{eqnarray}
    K_{22}^+ &=& \frac{1}{2} \big(  {}_{-2} Y_{22}(\iota, \varphi)
        + {}_{-2} Y_{2,-2}^* (\iota, \varphi)  \big)
        \nonumber \\
        &=& \frac{1}{4} \sqrt{\frac{5}{\pi}}
        \left( \cos^4 \frac{\iota}{2} + \sin^4 \frac{\iota}{2} \right) e^{\rmi 2\varphi}
        \nonumber \\       
        &=& \frac{1}{4} \sqrt{\frac{5}{\pi}} \frac{\left( 1+\cos^2\iota \right)}{2} e^{\rmi 2 \varphi},
        \\
        K_{22}^\times &=& \frac{\rmi}{2} \big(  {}_{-2} Y_{22}(\iota, \varphi)
        - {}_{-2} Y_{2,-2}^* (\iota, \varphi)  \big)
        \nonumber \\
        &=& \frac{\rmi}{4} \sqrt{\frac{5}{\pi}}
        \left( \cos^4 \frac{\iota}{2} - \sin^4 \frac{\iota}{2} \right) e^{\rmi 2\varphi}
        \nonumber \\
        &=& - \frac{\rmi}{4} \sqrt{\frac{5}{\pi}} \cos\iota e^{\rmi 2\varphi}.
    \end{eqnarray}
and so
\begin{equation}
    K_{2,-2}^+ = (K_{22}^+)^*,
    \qquad
    K_{2,-2}^\times = (K_{22}^\times)^*.
\end{equation}
Thus, one has
\begin{eqnarray}
    h_+ =& K_{22}^+ h_{22} + K_{2,-2}^+ h_{2,-2}
    = A_{22} \sqrt{\frac{5}{4\pi}} \frac{1 + \cos^2\iota}{2} \cos(\Phi_{22} - 2\varphi),
    \\
    h_\times =& K_{22}^\times h_{22} + K_{2,-2}^\times h_{2,-2}
    = A_{22} \sqrt{\frac{5}{4\pi}} \cos\iota \sin(\Phi_{22} - 2\varphi).
\end{eqnarray}

For non-precessing systems, Eq.~(\ref{eq:h_lm}) will be translate to
\begin{equation}
    \tdh_{\ell, -m} (f) = (-1)^\ell \tdh_{\ell m} (-f)^*
\end{equation}
in the Fourier domain.
For a given mode of \ac{gw} waveform, one has $h_{\ell m} \propto \exp[-\rmi m \phi_{orbit}]$,
where $\phi_{orbit}$ is the orbital phase of the \ac{gw} systems, and it always verifies with $\dot{\phi}_{orbit} >0$.
Thus, for non-precessing systems or in the processing frame for a binary with misaligned spins, an approximation often applied as
\begin{eqnarray}
    \tdh_{\ell m} (f) \simeq & 0 \quad {\rm for~ } m > 0, \quad f> 0, \\
    \tdh_{\ell m} (f) \simeq & 0 \quad {\rm for~ } m < 0, \quad f<0, \\
    \tdh_{\ell 0} (f) \simeq & 0.
\end{eqnarray}
In this way, for the positive frequencies $f >0$, $\tdh_{+, \times} = \sum_\ell \sum_{m<0} K^{+, \times}_{\ell, m} \tdh_{\ell m}$.

\paragraph{Eccentric mode decomposition}

Eccentric waveforms also generate the harmonics, which act similarly to higher modes but are described by the mean orbital frequency. Under the \ac{SPA}, there is a relationship between the mean orbital frequency $ F $ and the Fourier frequency $ f $ for different eccentric harmonics:
\begin{equation}
	f = j \cdot F(t_0).
\end{equation}
Here we use the index $j$ to distinguish eccentric harmonics from spin-weighted spherical harmonics above. $ t_0 $ is the time that gives the stationary point of $F$. The dominant eccentric harmonic is $ j=2 $.

With $ (\ell, m)=(2, 2) $, a frequency domain eccentric waveform can be written as
\begin{equation}
  \tilde{h}_{+,\times}=\sum_{j=1}^{10} \tilde{\mathcal{A}}_j \xi_{j}^{+,\times} e^{-{\rm i}\Psi_{j}}.
\end{equation}
Here
\begin{equation}
  \xi _j^{ + , \times } = C_{ + , \times }^{(j)} + {\rm i}S_{ + , \times }^{(j)},
\end{equation}
which is a function of $ (\iota, \varphi) $ and the eccentricity $ e(F) $. When $ e=0 $,
\begin{eqnarray}
        \xi _{j = 2}^ +  &= C_ + ^{(2)} + {\rm i}S_ + ^{(2)} = 4 \cdot \frac{{1 + {{\cos }^2}\iota }}{2}{e^{{\rm i} \cdot 2\varphi }}, \\
                \xi _{j = 2}^ \times  &= C_ \times ^{(2)} + {\rm i}S_ \times ^{(2)} = 4 \cdot \left( { - \cos \iota } \right){e^{{\rm i} \cdot 2\varphi }}, \\
		\xi _{j \ne 2}^{ + , \times } &= 0,
	\end{eqnarray}
which go back to the coefficients of the dominant mode $ (\ell, m)=(2, 2) $\cite{Yunes2009}. But for a non-zero eccentricity, one cannot explicitly write $ P_{\ell m} $ as we shown in Eq. (\ref{eq:plm}), and should directly use $P_+, P_\times$ in Eq. (\ref{eq:pppc}).

\subsection{Single arm response in time domain}

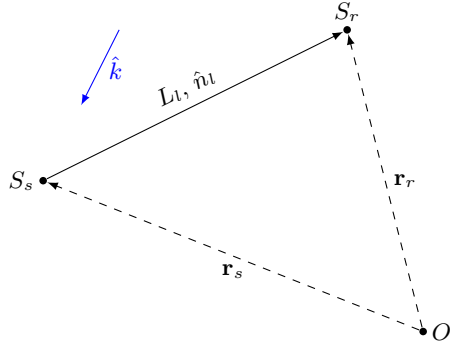
\begin{figure}[htbp]
    \centering
    \begin{tikzpicture}[>=latex]
        \node[circle,fill,inner sep=1pt] (O) at (0,2) {};
        \node[right] at (O) {$O$};
        \node[circle,fill,inner sep=1pt] (S1) at (-5,4) {};
        \node[left] at (S1) {$S_s$};
        \node[circle,fill,inner sep=1pt] (S2) at (-1,6) {};
        \node[above] at (S2) {$S_r$};

    \draw[->, dashed] (O) -- node[below] {$\bfr_s$} (S1);
    \draw[->, dashed] (O) -- node[right] {$\bfr_r$} (S2);

    \draw[->] (S1) -- node[sloped, above] {$L_l$, $\htn_l$} (S2);

    \draw[->, color=blue] (-4,6) -- node[right] {$\htk$} (-4.5,5);
    \end{tikzpicture}
    \caption{A radio signal send from the point $S_s$ travels along the arm $L_l$ in the direction of $\htn_l$ towards the receiver at $S_r$. 
    The coordinate origin is denoted by $O$, while point $S_s$ and point $S_r$ are located at $\bfr_s$ and $S_r$, respectively.}
    \label{fig:single_arm_resp}
\end{figure}

The effect of \acp{gw} on matter can be described as a tidal deformation.
To detect the \ac{gw}, one method is to test the distance changes between two spatially separated free-falling test masses.
Suppose that the photon travels along the direction of test mass 1 ($S_s$) to test mass 2 ($S_r$) as $\htn_l$, as shown in Fig.~\ref{fig:single_arm_resp}.
It follows a null geodesic, i.e., $ds^2 = 0$.
Thus, the metric reads
\begin{equation}
    cdt = \sqrt{(\delta_{ij} +h_{ij} (\xi) )  dx^i dx^j},
    \label{eq:cdt_xx}
\end{equation}
where
\begin{equation} 
    \xi(l) = t(l) - \htk \cdot \bfr (l)/c
    = t_s + l/c - \htk \cdot \left[ \bfr_s(t_s) + \htn (t_s) \, l \right]/ c,
\end{equation}
$\bfr_s$ is the position of $S_s$,
$\bfr(l)$ is the position of photon at time $t$,
$l = \sqrt{\sum_i (x^i -x_s^i)^2} = |\bfr(l) - \bfr_s|$,
$\htn = \frac{\bfr(l) - \bfr_s}{l}$,
and
\begin{eqnarray}
    \frac{d\xi}{dl/c} &=& 1 - \htk \cdot \htn_l(t_s),
    \quad
    \frac{dx^i}{dl} = \htn^i,
    \quad
    \htn^i \htn_i = 1
    \quad
    \nonumber \\
    & \Rightarrow &
    \nonumber \\
    \quad
    \frac{dx^i/c}{d\xi} &=& \frac{dx^i}{dl} \frac{dl/c}{d\xi}
    = \frac{\htn^i}{1 - \htk \cdot \htn}
\end{eqnarray}
With the above derivation, Eq.~(\ref{eq:cdt_xx}) can be rewritten as
\begin{eqnarray}
    dt &=& \sqrt{(\delta_{ij} + h_{ij}(\xi)) \frac{dx^i/c}{d\xi} \frac{dx^j/c}{d\xi}} d\xi
    = \sqrt{1 + h_{ij} \htn^i \htn^j} \frac{d\xi}{1 - \htk \cdot \htn}
    \nonumber \\
    &\approx & \left( 1 + \frac{1}{2} h_{ij} \htn^i \htn^j +\mathcal{O}(h^2) \right) \frac{d\xi}{1 - \htk \cdot \htn}.
    \end{eqnarray}

Then, from $S_s$ to $S_r$, the duration of the proper time will be
\begin{eqnarray}
    \int_{t_s}^{t_r} dt &=&\int_0^{L_l} \sqrt{ 1+ h_{ij} \htn^i \htn^j } \frac{d\xi}{1 - \htk \cdot \htn}
    \nonumber \\
        &\approx&  \int_0^{L_l}
        \left( 1+ \frac{1}{2} \htn_l^T \cdot \bfh \cdot \htn_l
        + \mathcal{O}(h^2) \right) \frac{d\xi}{1 - \htk \cdot \htn_l}
       \nonumber \\
        &=& \int_0^{L_l} dl/c + \int_0^{L_l} \frac{\htn_l^T \cdot \bfh \cdot \htn_l}{2(1-\htk \cdot \htn_l)} d\xi,
    \label{eq:integrate_tsr}
\end{eqnarray}
where $L_l$ is the length between $S_s$ and $S_r$,
$\htn_l$ is the unit vector of the photon propagation.


Here, if one supposes that the position of $S_s$ and $S_r$ does not change or changes very little during the photon moving from $S_s$ to $S_r$, which means $\htn_l(t_s) \approx \htn_l (t_r) = \htn_l$.
Then, for simplicity, the integral in Eq.~(\ref{eq:integrate_tsr}) can be rewritten as
\begin{equation}
    t_r = t_s + L_l/c + \frac{1}{2 ( 1 - \htk \cdot \htn_l)}
    \htn_l^T \cdot
    \left( \int_{\xi_s}^{\xi_r} \bfh  (\xi) d\xi \right)
    \cdot \htn_l.
    \label{eq:time_change}
\end{equation}
From this equation, one can directly get the path length fluctuations due to the \ac{gw}
\begin{equation}
    \delta l_{sr} (t) = \frac{c}{2 ( 1 - \htk \cdot \htn_l)}
    \htn_l^T \cdot
    \left( \int_{\xi_s}^{\xi_r} \bfh  (\xi) d\xi \right)
    \cdot \htn_l .
    \label{eq:length_change}
\end{equation}

Suppose the frequency of the photon is not changed during the photon's travel from $S_s$ to $S_r$.
Then the total phase change of the photon will be $\phi_{\rm tot} = 2\pi \nu_0 (t_r - t_s)$.
If there is no \ac{gw}, the phase change will be $\phi_{\rm ori} = 2\pi \nu_0 L/c$.
So, with the help of Eq.~(\ref{eq:time_change}), the phase fluctuations measured under the \ac{gw} will be
\begin{equation}
    \Delta \phi(t) = \phi_{\rm tot} - \phi_{\rm ori}
    = 2\pi \nu_0 \delta l_{sr}(t) / c.
    \label{eq:phase_chage}
\end{equation}

To get the time of reception changes concerning the time of emission, one can differentiate the above equation for $t_s$
\begin{eqnarray}
        \frac{dt_r}{dt_s} &=& 1 +
        \frac{1}{2 ( 1 - \htk \cdot \htn_l)}
        \htn_l^T \cdot
        \left( \int_{\xi_0}^{\xi_L} \frac{d\bfh  (\xi)}{d\xi}
        \frac{d\xi}{dt_s} d\xi \right)
        \cdot \htn_l
        \nonumber \\
        &=& 1 +
        \frac{1}{2 ( 1 - \htk \cdot \htn_l )}
        \htn_l^T \cdot
        \left[ \bfh(\xi_r) - \bfh(\xi_s) \right]
        \cdot \htn_l .
\end{eqnarray}
Here, we have used the assumption that the motion of $S_s$ and $S_r$ is much slower compared to the time of laser beam propagation, i.e., $d\bfr_s /dt_s \approx 0$ and $d\htn_l / dt_s \approx 0$, so $d\xi/dt_s = 1$.

The interferometers used to detect \acp{gw} do not emit single photons but continuous lasers with frequency $\nu(t)$.
If the phase change of the photon at $S_s$ and $S_r$ are the same, we have $d\phi/(2\pi) = \nu_s dt_s = \nu_r dt_r$.
Then one can get the dimensionless fractional frequency deviation $y^{GW}(t)$ as
\begin{eqnarray}
        y_{sr}^{GW}(t_r) &=& \frac{\nu_r - \nu_s}{\nu_s}
        = \frac{\nu_r}{\nu_s} - 1
        = \frac{d\phi/dt_r}{d\phi/dt_s} - 1
        \\
        &=& \frac{1}{ 1 +
        \frac{1}{2 ( 1 - \htk \cdot \htn_l )}
        \htn_l^T \cdot
        \left[ \bfh(\xi_r) - \bfh(\xi_s) \right]
        \cdot \htn_l } -1
        \\
        &\approx &
        \frac{1}{2 ( 1 - \htk \cdot \htn_l )}
        \htn_l^T \cdot
        \left[ \bfh(\xi_s) - \bfh(\xi_r) \right]
        \cdot \htn_l + \mathcal{O}(h^2).
    \label{eq:nu_chage}
\end{eqnarray}
Hence
\begin{eqnarray}
    \xi_s &=& t_s - \htk \cdot \bfr_s(t_s) /c
    \approx t_r - L_l/c - \htk \cdot [ \bfr_s(t_r - L_l/c) ]/c
    \nonumber    \\
    &\approx & t_r - L_l/c - \htk \cdot [\bfr_s(t_r) - \partial_t \bfr_s (t_r) L_l/c]/c
    \nonumber \\
    &\approx & t_r - L_l/c - \htk \cdot \bfr_s (t_r)/c,
    \\
    \xi_r &=& t_r - \htk \cdot \bfr_r (t_r)/c.
\end{eqnarray}
In the third line, we assumed that $\partial_t \bfr_s \ll c$.
Finally, one has the relative frequency deviation at the time of $t = t_r$ as \cite{Wahlquist:1987rx, armstrong_time-delay_1999, Katz:2022yqe}
\begin{equation}
    y_{slr}^{GW}(t) = \frac{1}{2 ( 1 - \htk \cdot \htn_l )} \,
        \htn_l^T \cdot
        \left[ \bfh(t - L_l/c - \htk \cdot \bfr_s/c)
        - \bfh(t - \htk \cdot \bfr_r/c) \right]
        \cdot \htn_l.
        \label{eq:freq_chage}
\end{equation}
When the photon reflected from $S_r$ to $S_s$, we have
\begin{equation}
    y_{rls}^{GW}(t) = \frac{1}{2 ( 1 + \htk \cdot \htn_l)} \,
        \htn_l^T \cdot
        \left[ \bfh(t - L_l/c - \htk \cdot \bfr_r / c)
        - \bfh(t - \htk \cdot \bfr_s /c) \right]
        \cdot \htn_l .
\end{equation}
Considering that the \ac{gw} is described in the SSB coordinate, thus, one can redefine some parameters as
\begin{equation}
    y_{slr}^{\rm GW} (t) =
    \frac{1}{2(1 - \htk \cdot \htn_l)}
    \left[ H(t - L/c -\htk\cdot \bfr_s/c) - H(t - \htk \cdot \bfr_r/c) \right],
\end{equation}
where
\begin{eqnarray}
    H(t) &=& n_l^i h_{ij}(t) n_l^j
        = n_l^i (h_+ \epsilon_{ij}^+ + h_\times \epsilon_{ij}^\times ) n_l^j
        \nonumber \\
        &=& n_l^i \left[ h_+ (u_i u_j -v_i v_j) + h_\times (u_i v_j + v_i u_j) \right] n_l^j
        \nonumber \\
        &=& h_+ (n_l^i u_i u_j n_l^j - n_l^i v_i v_j n_l^j)
        + h_\times (n_l^i u_i v_j n_l^j + n_l^i v_i u_j n_l^j)
        \nonumber \\
        &=& h_+ \left[ (\htn_l \cdot \hat{u})^2 - (\htn_l \cdot \hat{v})^2 \right]
        + h_\times \cdot 2 (\htn_l \cdot \hat{u}) (\htn_l \cdot \hat{v} )
        \nonumber \\
        &=& h_+ \zeta_l^+ + h_\times \zeta_l^\times,
    \end{eqnarray}
and
\begin{eqnarray}
    \zeta_l^+ =& \htn_l \cdot \epsilon^+ \cdot \htn_l
    = (\htn_l \cdot \hat{u})^2 - (\htn_l \cdot \hat{v})^2,
    \\
    \zeta_l^\times =& \htn_l \cdot \epsilon^\times \cdot \htn_l
    = 2 (\htn_l \cdot \hat{u}) (\htn_l \cdot \hat{v}).
\end{eqnarray}

For the two-way response, one can get \cite{estabrook1975response}
\begin{eqnarray}
    y_{sls}^{GW}(t) &= \, & \frac{\nu}{\nu_0} - 1 = \frac{\nu}{\nu'}\frac{\nu'}{\nu_0} - 1
        = \big[ y_{slr}^{GW} (t - L_l/c) + 1 \big] \big[y_{rls}^{GW} (t) +1 \big] - 1
        \nonumber \\
        &\approx \, & y_{slr}^{GW} (t-L/c) + y_{rls}^{GW} (t) + \mathcal{O}(h^2)
        \nonumber \\
        &= \, & \frac{1}{2} \bigg\{ (1 + \htk \cdot \htn) \big[ \Psi_l(t - 2L_l/c) - \Psi_l(t- L/c) \big]
        \nonumber \\
        && + (1 - \htk \cdot \htn) \big[ \Psi_l(t - L_l/c) - \Psi_l(t) \bigg\}
        \nonumber \\
        &= \, & \frac{1 + \htk \cdot \htn}{2} \Psi_l (t - 2L_l/c)
        - \htk \cdot \htn \, \Psi_l (t - L_l/c)
        - \frac{1 - \htk \cdot \htn}{2} \Psi_l(t),
    \end{eqnarray}
where (using $\htn = \htn(t)$ for convenient)
\begin{equation}
    \Psi_l (t') = \frac{\htn_l^T \cdot \bfh(t' - \htk \cdot \bfr(t')/c) \cdot \htn_l}{1 - (\htk \cdot \htn_l)^2}.
\end{equation}
However, one should note that the above derivation is based on the assumption that the positions of the spacecraft change very little between the photon sent from $S_s$ to $S_r$.

\subsection{Single arm response in frequency domain}

Adopting the Fourier transform, the \ac{gw} in the frequency domain will be \cite{Romano:2016dpx}
\begin{equation}
    h_0(t, \vec{x}) = h(t - d(t) )
    = \int df e^{\rmi 2\pi f (t - d(t))} \tilde{h}(f)
    {\rm or~ }
    h(\xi) = \int df e^{\rmi 2\pi f\xi} \tdh(f).
\end{equation}
With the Fourier transform, the path length fluctuations could be rewritten as
\begin{eqnarray}
        \delta l_{sr} (t) &=& \frac{c}{2 ( 1 - \htk \cdot \htn_l )}
        \htn_l^T \cdot
        \left( \int_{\xi_s}^{\xi_r} d\xi \int df \, e^{\rmi 2\pi f \xi} \, \tbf{h} (f) \right)
        \cdot \htn_l
        \nonumber \\
        &= & \frac{c}{2 ( 1 - \htk \cdot \htn_l)}
        \htn_l^T  \cdot
        \left( \int df \,
        \left(
        e^{\rmi 2\pi f \xi_r}
        - e^{\rmi 2\pi f \xi_s}
        \right)
        \frac{\tbf{h} (f)}{\rmi 2\pi f} \right)
        \cdot \htn_l
        \nonumber \\
        &=& \, L_l \,
        \htn_l^T \cdot
        \left( \int df \, e^{\rmi 2\pi f t}
        \mathcal{T}_{sr} (\htk, f, t) \tbf{h}(f) \right)
        \cdot \htn_l
    \label{eq:length_change_FD}
\end{eqnarray}
where $\mathcal{T}_{sr} (\htk, f, t)$ is the transfer function \cite{Romano:2016dpx}
\begin{eqnarray}
    \mathcal{T}_{sr}(f, t) &=&
        \frac{c / L}{2 ( 1 - \htk \cdot \htn_l)}  \,
        \frac{1}{\rmi 2\pi f}
        \left(
        e^{-\rmi 2\pi f \htk \cdot \bfr_r/c}
        - e^{-\rmi 2\pi f (L_l + \htk \cdot \bfr_s)/c}
        \right)
        \nonumber \\
        &=& \frac{c / L}{2 ( 1 - \htk \cdot \htn_l)}  \,
        \frac{1}{\rmi 2\pi f}
        \left(
        e^{\rmi \pi f [L - \htk \cdot (\bfr_r - \bfr_s)]/c}
        - e^{-\rmi \pi f [L_l - \htk \cdot (\bfr_r - \bfr_s) ]/c}
        \right)
        \nonumber\\
        && e^{-\rmi \pi f [L + \htk \cdot (\bfr_r + \bfr_s)]/c}
        \nonumber \\
        &= & \frac{c / L}{2 ( 1 - \htk \cdot \htn)}  \,
        \frac{\rmi 2\sin\left(
        \pi f L/c (1 - \htk \cdot \htn)
        \right)}{\rmi 2\pi f}
        e^{-\rmi \pi f [L + \htk \cdot (\bfr_r + \bfr_s)]/c}
        \nonumber \\
        &= & \frac{1}{2} {\rm sinc}
        \left( \pi f L/c (1 - \htk \cdot \htn)   \right)
        \exp\left\{-\rmi \pi f [L + \htk \cdot (\bfr_r + \bfr_s)]/c \right\}.
    \end{eqnarray}
Finally, one can define the one-arm detector tensor as
\begin{equation}
    \bfD(\htk, f, t) = \htn (t) \otimes \htn(t) \,
    \mathcal{T}(\htk, f, t),
\end{equation}
and the path length fluctuation in the frequency domain will be
\begin{equation}
    \frac{\delta \tdl_{sr} (f)}{L_l} = \bfD(\htk, f, t) \,\mathbf{:}\, \tbf{h} (f),
\end{equation}
where $(\htn \otimes \htn)_{ij} = \htn_i \htn_j$, $\bfA \mathbf{:} \bfB = A_i B_i$.
Similarly, the phase fluctuation in the frequency domain will be
\begin{equation}
    \Delta \tilde{\phi}(f) = \frac{2 \pi \nu_0}{c} \delta \tdl_{sr}(f).
\end{equation}

Similarly, one can derive the relative frequency derivation in the frequency domain with the Fourier transform, which reads
\begin{eqnarray}\label{eq:yslr}
        y_{slr}^{GW} (t)
        &=& \frac{1}{2 ( 1 - \htk \cdot \htn_l)} \,
        \htn_l^T \cdot
        \bigg[
        \int df' e^{\rmi 2\pi f' (t - L_l/c - \htk \cdot \bfr_s/c)}
        \bfh(f')
        \nonumber \\
        && - \int df'' e^{\rmi 2\pi f'' (t - \htk \cdot \bfr_r/c)}
        \bfh(f'') \bigg]
        \cdot \htn_l
        \nonumber \\
        &=& \frac{1}{2 ( 1 - \htk \cdot \htn_l)} \,
        \htn_l^T \cdot
        \int df\, e^{\rmi 2\pi f t}\, \tbf{h}(f) \left[
        e^{-\rmi 2\pi f (L_l + \htk \cdot \bfr_s)/c}
        - e^{-\rmi 2\pi f \, \htk \cdot \bfr_r/c}
        \right]
        \cdot \htn_l
        \nonumber \\
        &=& \int df \, e^{\rmi 2\pi ft} \,
        \left[
            e^{-\rmi 2\pi f (L_l + \htk \cdot \bfr_s)/c}
            - e^{-\rmi 2\pi f \, \htk \cdot \bfr_r/c}
        \right]
        \frac{\htn_l^T  \cdot  \tbf{h}(f) \cdot \htn_l}
        {2 ( 1 - \htk \cdot \htn_l )}
        \nonumber \\
        &=& \int df \, e^{\rmi 2\pi ft} \,
        \left[
            e^{-\rmi \pi f [L_l + \htk \cdot (\bfr_s - \bfr_r)]/c}
            - e^{-\rmi \pi f ( \htk \cdot \bfr_r - L_l- \htk\cdot \bfr_s)/c}
        \right]
    \nonumber \\
    &&
        e^{-\rmi \pi f [L_l + \htk \cdot (\bfr_s + \bfr_r)]/c}
        \,
        \frac{\htn_l^T  \cdot  \tbf{h}(f) \cdot \htn_l}
        {2 ( 1 - \htk \cdot \htn_l )}
        \nonumber \\
        &=& \int df \, e^{\rmi 2\pi ft} \,
        \bigg\{
            - \frac{\rmi \sin\left[ \pi f L_l/c ( 1- \htk\cdot \htn_l) \right]}
            { ( 1 - \htk \cdot \htn_l )} \,
            e^{-\rmi \pi f [L_l + \htk \cdot (\bfr_s + \bfr_r)]/c}
            \nonumber \\
            && \qquad \qquad\qquad
            \htn_l^T  \cdot  \tbf{h}(f) \cdot \htn_l
        \bigg\}
        \nonumber \\
        &=& \int df \, e^{\rmi 2\pi ft} \,
        \bigg\{
            - \frac{\rmi\pi fL_l}{ c}
            \mathrm{sinc}\left[ \pi f L_l/c ( 1- \htk\cdot \htn_l) \right] \,
            \nonumber \\
            && e^{-\rmi \pi f [L_l + \htk \cdot (\bfr_s + \bfr_r)]/c}
            \,
            \htn_l^T \cdot \tbf{h}(f) \cdot \htn_l
        \bigg\}
        \nonumber \\
        &=& \int df \, e^{\rmi 2\pi ft} \,
        \left\{ 
            - \frac{\rmi 2\pi f L_l}{c} \mathcal{T}_{sr} (f,t) \,            
            (h_+ \zeta_l^+ + h_\times \zeta_l^\times)
        \right\}.
\end{eqnarray}

Here  $\htn_l \cdot \tbf{h}(f) \cdot \htn_l
= \htn_l \cdot (\tdh_+ \epsilon^+ + \tdh_\times \epsilon^\times ) \cdot \htn_l
= \tdh_+ \zeta_l^+ + \tdh_\times \zeta_l^\times$.
If the \ac{gw} tensor can be decomposed into $\tbf{h} = \mathbf{P}(f) \tilde{h}(f)$, where $\bfP = e^+ + e^\times$.
Thus, the transfer function for the relative frequency deviation can be written as \cite{Marsat:2018oam, Marsat:2020rtl}
\begin{eqnarray}
    G_{slr}^{GW} (f, t) &=& -\frac{\rmi\pi fL_l}{c}
        \mathrm{sinc}\left[ \pi f L_l/c ( 1- \htk\cdot \htn_l) \right] \,
        \nonumber \\
        && e^{-\rmi \pi f [L + \htk \cdot (\bfr_s + \bfr_r)]/c}
        \,
        \htn_l^T \cdot  \mathbf{P}(f) \cdot \htn_l.
\end{eqnarray}
For the multiple modes, the transfer function $G_{slr}^{\ell m}(f,t)$ has the same form as the above equation, where just $\bfr$ should be changed to $P_{\ell m}$, i.e.,
\begin{equation}\label{eq:gslr}
    G_{slr}^{\ell m} (f, t) = -\frac{\rmi\pi fL_l}{ c}
        \mathrm{sinc}\left[ \pi f L_l/c ( 1- \htk\cdot \htn_l) \right] \,
        e^{-\rmi \pi f [L + \htk \cdot (\bfr_s + \bfr_r)]/c}
        \,
        \htn_l^T \cdot  P_{\ell m} \cdot \htn_l .
\end{equation}
With help of $\xi_+$ and $\xi_\times$, the part of $\htn_l \cdot P_{\ell m} \cdot \htn_l$ will be
\begin{eqnarray}
    \htn_l \cdot P_{\ell m} \cdot \htn_l &= &
            \frac{1}{2} {}_{-2} Y_{\ell m} (\iota, \varphi)
            e^{- \rmi 2 \psi} (\zeta_l^+ + \rmi \zeta_l^\times)
            \nonumber \\
            && + \frac{1}{2} (-1)^{\ell} {}_{-2} Y_{\ell, -m}^* (\ell, \varphi)
            e^{+\rmi 2\psi} (\xi_l^+ - \rmi \xi_l^\times)
            .
    \end{eqnarray}
One should note that in the previous equations, when the higher modes are considered, the time-frequency relationship should be considered.
With the help of the stationary phase approximation, the time-frequency relationship will be
\begin{equation}
    t_f^{\ell m} = - \frac{1}{2\pi} \frac{d\Psi_{\ell m}}{df},
\end{equation}
for different modes.

As for \ac{GW} with eccentricity, we could not simply calculate the response function using the formulae above, even if it only has the dominant spin-weighted spherical harmonic $ (\ell, m)=(2, 2) $. Different eccentric harmonics also have different time-frequency correspondence, so we need to write \cite{Wang2023}
\begin{equation}
	t_f^j = \frac{1}{{2\pi }}\frac{{d{\Psi _j}}}{{df}}.
\end{equation}
Then we decompose $\tbf{h}$ into eccentric harmonics $\tbf{h}_j$, i.e.
\begin{eqnarray}
		\tbf{h} &=\sum\limits_{j} \tbf{h}_j,\\
		{\tbf{h}}_j &={{P_ + }\tilde h_j^ +  + {P_ \times }\tilde h_j^ \times },
	\end{eqnarray}
and rewrite Eq. (\ref{eq:yslr})-(\ref{eq:gslr}):
\begin{eqnarray}
    {{\tilde y}_{slr}} &=& \sum\limits_j {\mathcal{T}_{slr}^j(f):{\tilde h}_j},
\\
    \mathcal{T}_{slr}^j(f) &=& {G_{slr}}\left( {f,t_f^j} \right),
\\
      G_{slr}^{\ell m} (f, t) &=& -\frac{\rmi\pi fL_l}{ c} \mathrm{sinc}\left[ \pi f L_l/c ( 1- \htk\cdot \htn_l) \right] \, 
      \nonumber \\
      && e^{-\rmi \pi f [L + \htk \cdot (\bfr_s + \bfr_r)]/c} \, \htn_l^T \otimes \htn_l .
\end{eqnarray}

\subsection{Response for the mildly chirping signals}
\label{sec:midly_chirp}

For mildly chirping binary sources that do not contain the Fourier integral,
one can assume that the phase of the \ac{gw} can be approximated as \cite{Cornish:2002rt}.
\begin{equation}
    \Phi(\xi) = 2\pi f_0 \xi + \pi \dot{f}_0 \xi^2 + \varphi_0,
\end{equation}
where $f_0, \dot{f}_0$ and $\varphi_0$ are the initial frequency, frequency deviation and phase, respectively.
Thus, the instantaneous frequency can be given as
\begin{equation}
    \frac{1}{2\pi} \frac{\partial \Phi(\xi) }{\partial t}
    = \frac{1}{2\pi} \frac{\partial \Phi(\xi)}{\partial \xi} \frac{\partial \xi}{\partial t}
    = (f_0 + \dot{f}_0 \xi)
    \left( 1 - \htk \cdot \frac{\partial \bfr (t)}{\partial t} \right).
\end{equation}
According to the equation, we may assume a fixed frequency at $\xi_0$ as
\begin{equation}
    f_s = f_0 + \dot{f}_0 \xi_0,
\end{equation}
and the index $s$ denotes the dependency of the approximated frequency on the time of emission $\xi_0$.
Here, assuming that the frequency of the \ac{gw} changes very little, i.e., $\dot{f}_0 (\xi_L - \xi_0) \ll f_0$.
Then
\begin{equation}
    \Phi(\xi) \approx \int dt \, 2\pi f_s
    \left( 1 - \htk \cdot \frac{\partial \bfr (t)}{\partial t} \right)
    = \int d\xi \, 2\pi f_s
    =2\pi (f_0 + \dot{f}_0 \xi_0 ) \xi + C,
\end{equation}
where $C$ is some integration constant.
Meanwhile, the amplitude of the wave also changes little.
Then the plane wave can be described as
\begin{equation}
    h(\xi) = A(\xi) e^{\rmi 2\pi f_s \xi}
    \approx A(\xi_0) e^{\rmi 2\pi f_s \xi_0} e^{\rmi 2\pi f_s (\xi -\xi_0)}
    = h(\xi_0) e^{\rmi 2\pi f_s (\xi - \xi_0)}.
\end{equation}
In this way, the integration of the \ac{gw} tensor fluctuation will be \cite{Cornish:2002rt}
\begin{eqnarray}
        \int_{\xi_0}^{\xi_L} \bfh(\xi) d\xi
        &=& \bfP \int h(\xi_0) e^{\rmi 2\pi f_s (\xi - \xi_0)} d\xi
        = \bfP \frac{1}{\rmi 2\pi f_s} \left( h(\xi_L) - h(\xi_0) \right)
        \nonumber \\
        &=& \bfP \frac{1}{\rmi 2\pi f_s} h(\xi_0)
        \left( e^{\rmi 2\pi f_s (\xi_L - \xi_0)} - 1 \right)
        \nonumber \\
        &=& \bfP \frac{\sin \left[ \pi f_s (\xi_L - \xi_0) \right] }{\pi f_s}
        e^{\rmi \pi f_s (\xi_L - \xi_0)} h(\xi_0)
        \nonumber \\
        &=& \bfP \frac{\sin \left[ \pi f_s (\xi_L - \xi_0) \right] }{\pi f_s}
        e^{\rmi \pi f_s (\xi_L + \xi_0)} A(\xi_0)
        \nonumber \\
        &= &\bfP \frac{(1 -\htk \cdot \htn) L}{c}
        {\rm sinc} \left[ \frac{\pi f_s L}{c} \, (1 - \htk \cdot \htn) \right]
        \nonumber \\
        && e^{ - \rmi \pi f_s [L + \htk \cdot (\bfr_r + \bfr_s)]/c}
        A(\xi_0) e^{\rmi 2\pi f_s t_r}
        \nonumber \\
        &=& 2\, \bfP \frac{(1 -\htk \cdot \htn) L}{c}
        \mathcal{T}_{sr} (\htk, f_s, t_r)
        A(\xi_0) e^{\rmi 2\pi f_s t_r},
    \end{eqnarray}
where $\bfP$ is the unit tensor matrix of \ac{gw}.
Here we have used
\begin{eqnarray}
    \xi_L - \xi_0 &=& (t_r -\htk \cdot \bfr_r) - (t_s - \htk \cdot \bfr_s)
        = (1 - \htk \cdot \htn) L/c ,
        \\
        \xi_L + \xi_0 &=& (t_r -\htk \cdot \bfr_r/c) + (t_s - \htk \cdot \bfr_s/c)
        \nonumber \\
        &\approx& 2 t_r - L/c - \htk \cdot (\bfr_s + \bfr_r)/c.
    \end{eqnarray}
If the amplitude of \ac{gw} is some constant, then the path length variation defined in Eq.~(\ref{eq:length_change}) will be
\begin{equation}
    \frac{\delta l_{sr}}{L} (t)
    \approx \mathcal{T}_{sr}(\htk, f_s, t) \,
    \htn \cdot \bfh(t) \cdot \htn.
\end{equation}
And according to Eq.~(\ref{eq:nu_chage}), one can find that
\begin{equation}
    \frac{\delta \nu}{\nu_0} = -\frac{\rmi 2\pi f_s L}{c} \frac{\delta l}{L}.
\end{equation}
This is similar to the response in the frequency domain, and one should note that the above formula is valid only when the \ac{gw} is some mildly chirping signals or some monochromatic signals.

\subsection{Response to the Stochastic backgrounds}
\label{sec:orf}

There are many \ac{gw} events in the Universe.
And the superposition of weak \acp{gw} can form an \ac{sgwb}.
It can be expanded in terms of plane waves
\begin{eqnarray}
        h_{ij} (t, \bfr) 
        = & \sum_{+,\times} \int \hat{\Omega}_\htk \int df \,
        \tdh_{+,\times} (f, \htk) \, \epsilon^{+,\times} (\htk) : \bfD (\htk, f)
        e^{-\rmi 2\pi f (t - \htk \cdot \bfr/c)} .
    \label{eq:sgwb_h}
\end{eqnarray}
where $\int d\hat{\Omega} = \int_0^{2\pi} d\phi \int_0^{\pi} \sin \theta d\theta$ denotes an integral in all the sky,
$\bfr$ is the location of the measurement at time $t$.

The detector response to the superposition of \acf{gw} can be written as the convolution of the metric perturbations $h_{ij} (t, \bfr)$ with the impulse response $F^{ij} (t, \bfr)$ of the detector
\begin{eqnarray}
        s(t) &=& (\bfR \star \bfh) (t, \htr)
        \equiv \int d\tau \int d^3 x F^{ij} (\tau, \bfx) h_{ij} (t-\tau, \bfr -\bfx)
        \nonumber \\
        &=& \int df \int d \hat{\Omega}_\htk \, R^{ij} (f, \htk)
        h_{ij} (f, \htk) e^{-\rmi 2\pi ft}
        = \int df \tds (f) e^{-\rmi 2\pi f t},
    \label{eq:sgwb_rh}
\end{eqnarray}
where
\begin{equation}
    R^{ij} (f, \htk) =
    e^{\rmi 2\pi f \htk \cdot \bfr /c}
    \int d\tau \int d^3 x F^{ij} (\tau, \bfx)
    e^{- \rmi 2\pi f (\tau + \htk \cdot \bfx /c ) }.
\end{equation}
is the response function in the frequency domain,
and $\tds(f)$ is the signal in frequency domain.

For individual detectors, the analysis focuses on the response function of a single channel, particularly the orthogonal TDI channels such as the A/E channels. However, when considering detector networks, the correlation between different channels becomes a critical factor.
In this context, the response function generalizes to the overlap reduction function (ORF) for distinct channels. 
Notably, the ORF exhibits a time dependence as a result of the orbital motion of the detectors. 
To account for this variability, one can define the time-averaged ORF over the total correlation time $T_{\rm tot}$ as follows~\cite{Liang:2021bde}:
\begin{equation}
\bar{\Gamma}_{ij}(f)=
\sqrt{\frac{1}{T_{\rm tot}}\int_{0}^{T_{\rm tot}}{\rm d}t\,|\hat{\Gamma}_{ij}(f,t_{0})|^{2}}.
\end{equation}

For networks such as TianQin I+II or TianQin + LISA, four distinct channel pairs can be utilized for cross-correlation analysis. 
Consequently, the total ORF for these networks is defined as the quadrature sum of the individual time-averaged ORFs~\cite{Seto:2020mfd}
\begin{equation}
\label{eq:gamma_total}
\Gamma_{{\rm tot}}(f)=
\sqrt{\sum_{I,J}\big|\bar{\Gamma}_{IJ}(f)\big|^{2}},
\end{equation}
where $I$ and $J$ label a pair of orthogonal channels. 

\section{Time Delay Interference}
\label{sec:TDI}




The signal transmitted from spacecraft $S_s$ that is received at spacecraft $S_r$ at time $t_r$ has its phase compared to the local reference to give the output of the phase change $\Phi_{sr} (t_r)$ \cite{Cornish:2002rt}.
The phase difference has contributions from the
laser phase noise $C(t)$,
optical path length variations,
shot noise $n^s(t)$
and acceleration noise $\mathbf{n}^a(t)$ \cite{Cornish:2002rt}
\begin{eqnarray}
    \Phi_{slr} (t_r) &=& \cboxed{C_s(t_s) - C_r(t_r)}
    + 2\pi \nu_0 (\cboxed[cyan]{\delta l_l (t_s)} + \cboxed[blue]{\Delta l_l (t_s)})
    \nonumber \\
    && + \cboxed[green]{n^s_{sr} (t_r)}
    - \htn_l (t_s) \cdot \left[ \cboxed[yellow]{\mathbf{n}^a_{sr}(t_r) - \mathbf{n}^a_{rs}(t_s) } \right],
    \label{eq:detect_phase}
\end{eqnarray}
where $t_s$ is given implicitly by $t_s = t_r - \ell_{sr}(t_s)$
and $\nu_0$ is the laser frequency.
The optical path length variations caused by gravitational waves are $\delta l_l(t_s)$,
and those caused by orbital effects is $\Delta l_l (t_s)$.
From Eq.~(\ref{eq:detect_phase}), one can find that the space-based \ac{gw} detection suffers from laser phase noise, which can be alleviated through \ac{tdi} technology.
\ac{tdi} involves heterodyne interferometry with unequal arm lengths and independent phase-difference readouts \cite{Tinto:2020fcc}.
By essentially constructing a virtually equal-arm interferometer, the laser phase noise cancels out exactly.

\subsection{General \ac{tdi} combination}

\begin{figure}[htpb]
    \centering
    \begin{tikzpicture}[>=latex]
        \node (S1) at (-3,0) {S1};
        \node (S1i) at (-2.7,0.17) {};
        \node (S2) at (0,5.2) {S2};
        \node (S2i) at (0,4.85) {};
        \node (S3) at (3,0) {S3};
        \node (S3i) at (2.7,0.17) {};

        \foreach \la in {S1, S2, S3}
        \draw[line width=1pt] (\la) circle[radius=0.3];

        \foreach \li/\lo/\ii/\ts in {S1i/S2i/3/below, S2i/S3i/1/below, S3i/S1i/2/above}
        \draw[color=red,->,line width=2pt] (\lo) -- (\li) %
        node[midway,sloped,\ts,color=black] {$L_\ii$} 
        node[near end,sloped,\ts,color=black] {$\htn_\ii$};

        \foreach \li/\lo/\ii/\ts in {S1/S2/3/above, S2/S3/1/above, S3/S1/2/below}
        \draw[color=blue,<-,dashed,line width=2pt] (\lo) -- (\li) %
        node[midway,sloped,\ts,color=black] {$L'_\ii$} %
        ;
    \end{tikzpicture}
    \caption{Illustration of detector constellation. Three satellites are marked as 1, 2, and 3. Laser paths are marked as $L_i$ and $L'_i$, where $L'_i$ represents the direction opposite to $L_i$. The direction of unit vector $\htn_i$ is the same as that of $L_i$.}
    \label{fig:TDIlinks}
\end{figure}
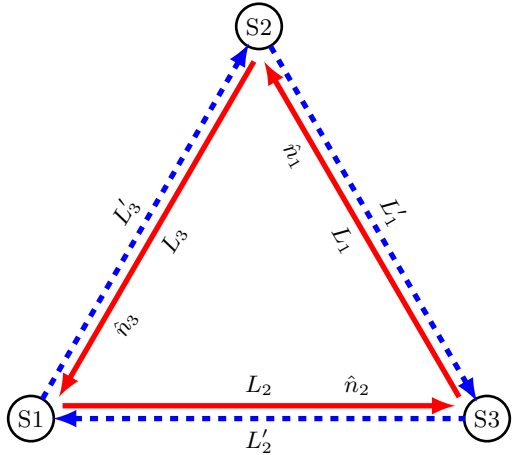

Before introducing the \ac{tdi}, let's first introduce some definitions.
In Fig.~\ref{fig:TDIlinks}, the satellite numbers in space are defined clockwise.
The definition of the laser path counterclockwise is the positive direction ($\htn_i$), denoted as $L_i$, and clockwise is the negative direction ($-\htn_i$), denoted as $L'_i$.
The arm length $|L_i|$ is defined as the distance between the other two satellites facing the satellite $i$, where $i = 1,2,3$.

As shown in Fig.~\ref{fig:TDIlinks}, let $\vec{X}_i$ as the $i$-th spacecraft, $l_{ij}$ is the distance between the $i$-th and $j$-th spacecrafts, then
\begin{eqnarray}
    L_1 &=& \vec{u}_{32} \, l_{32} = \vec X_2 - \vec X_3 ,\, 
    L_2 = \vec{u}_{13} \, l_{13} = \vec X_3 - \vec X_1, \, 
    \nonumber \\
    L_3 &=& \vec{u}_{21} \, l_{21} = \vec X_1 - \vec X_2.
    \\
    \htn_1 &=& \vec{u}_{32} = \frac{\vec X_2 -\vec X_3}{|\vec X_2 - \vec X_3|} ,\, 
    \htn_2 = \vec{u}_{13} = \frac{\vec X_3 -\vec X_1}{|\vec X_3 - \vec X_1|} ,\, 
    \nonumber \\
    \htn_3 &=& \vec{u}_{21} = \frac{\vec X_1 -\vec X_2}{|\vec X_1 - \vec X_2|}
\end{eqnarray}
Let $s_1$ as the time-dependent phase change signals received by the $1$-st spacecraft, which is sent from the $2$-nd spacecraft and propagates along the link $L_3$.
One can also sign it as $s_{231}$.
Similarly, let $s'_1$ as the signal received by spacecraft $1$, which is sent from spacecraft $3$ and propagates along $L'_2$, or recorded as $s_{321}$.
As shown in Fig.~\ref{fig:TDIlinks}, there are six independent laser links.

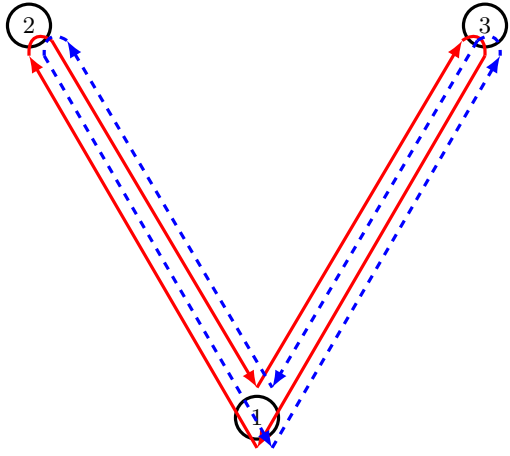
\begin{figure}
    \centering
    \begin{tikzpicture}[very thick,>=latex]
        \node[draw, circle] (s1) at (0,-5.2) {1};
        \node[draw, circle] (s2) at (-3,0) {2};
        \node[draw, circle] (s3) at (3,0) {3};

        \draw[color=red, ->] (0,-5.6) -- (-3,-0.4);
        \draw[color=red, ->] (-2.7,-0.2) -- (0,-4.8);
        \draw[color=red,rounded corners=10pt] (-2.7,-0.2) --(-3,0) -- (-3,-0.4);
        \draw[color=red, <-] (0,-5.6) -- (3,-0.4);
        \draw[color=red, <-] (2.7,-0.2) -- (0,-4.8);
        \draw[color=red, rounded corners=10pt] (2.7,-0.2) --(3,0) -- (3,-0.4);

        \draw[color=blue, dashed, <-] (0.2,-5.6) -- (-2.8,-0.4);
        \draw[color=blue, dashed, <-] (-2.5,-0.2) -- (0.2,-4.8);
        \draw[color=blue, dashed, rounded corners=10pt] (-2.5,-0.2) --(-2.8,0) -- (-2.8,-0.4);
        \draw[color=blue, dashed, ->] (0.2,-5.6) -- (3.2,-0.4);
        \draw[color=blue, dashed, ->] (2.9,-0.2) -- (0.2,-4.8);
        \draw[color=blue, dashed, rounded corners=10pt] (2.9,-0.2) --(3.2,0) -- (3.2,-0.4);
    \end{tikzpicture}
    \caption{Michelson-like TDI-X channel of first generation \ac{tdi}.}
    \label{fig:TDI-X-1.0}
\end{figure}

The first generation \ac{tdi} combination does not consider the rotation and flexing of the spacecraft constellation, which is only valid for a static constellation, i.e.,
\begin{equation}
    L_i(t) = L_i = {\rm const},
    \qquad
    L_i = L_{i'}.
\end{equation}
This means that all the arm lengths remain constant as time evolves,
and the time duration of photon propagation along the arm is independent of the direction of the photons.
The 1.5 or modified \ac{tdi} generation is valid for a rigid but rotating spacecraft constellation, i.e.,
\begin{equation}
    L_i(t) = L_i = {\rm const},
    \qquad
    L_i \neq L_{i'}.
\end{equation}
The propagation direction of photons should be considered.
The second generation \ac{tdi} combination is applied to consider a rotating and flexing constellation, i.e.,
\begin{equation}
    L_i(t) = L_i + \dot{L}_i \, t,
    \qquad
    L_i \neq L_{i'}.
\end{equation}
The arm length changes linearly in time, and relative to the velocity of $\dot{L}_i$.
Here, define the time delay operator as $\mcD_i$, where
\begin{eqnarray}
    \mcD_i  x(t) & \equiv& x(t - L_i/c), \\
    \mcD_i \mcD_j x(t) &=& \mcD_{ij} \, x(t) \equiv x(t - L_i/c -L_j/c).
\end{eqnarray}
Then one can define the 1.5 generation unequal arm Michelson-like combination as (see Fig.~\ref{fig:TDI-X-1.0}) \cite{armstrong_time-delay_1999}
\begin{eqnarray}
    X_{1.5} &=& y_{32'1} + \mcD_{2'} \left[ y_{123}
        + \mcD_{2} \left( y_{231} + \mcD_{3} y_{13'2} \right) \right]
        \nonumber \\
        && - y_{231} - \mcD_{3} \left[ y_{13'2}
        + \mcD_{3'} \left( y_{32'1} + \mcD_{2'} y_{123} \right) \right]
        \nonumber \\
        &= & y_{32'1} + \mcD_{2'} y_{123} + \mcD_{2'2} y_{231} + \mcD_{2'23} y_{13'2}
        \nonumber \\
        && - y_{231} - \mcD_{3} y_{13'2} - \mcD_{33'} y_{32'1} - \mcD_{33'2'} y_{123}.
\end{eqnarray}
For the generation 2.0, one has \cite{armstrong_time-delay_1999}
\begin{eqnarray}
        X_{2.0} =& \, y_{32'1} + \mcD_{2'} y_{123}
        + \mcD_{2'2} y_{231} + \mcD_{2'23} y_{13'2}
        \nonumber \\
        & + \mcD_{2'233'} y_{231} + \mcD_{2'233'3} y_{13'2}
        + \mcD_{2'233'33'} y_{32'1} + \mcD_{2'233'33'2'} y_{123}
        \nonumber \\
        & - y_{231} - \mcD_{3} y_{13'2}
        - \mcD_{33'} y_{32'1} - \mcD_{33'2'} y_{123}
        \nonumber \\
        & - \mcD_{33'2'2} y_{32'1} - \mcD_{33'2'22'} y_{123}
        - \mcD_{33'2'22'2} y_{231} - \mcD_{33'2'22'23} y_{13'2}.
\end{eqnarray}
The $Y$ and $Z$ channels can be generated by cyclic permutation of indices: $1 \rightarrow 2 \rightarrow 3 \rightarrow 1$.

\begin{figure}[htbp]
\centering
    \includegraphics[width=0.3\linewidth]{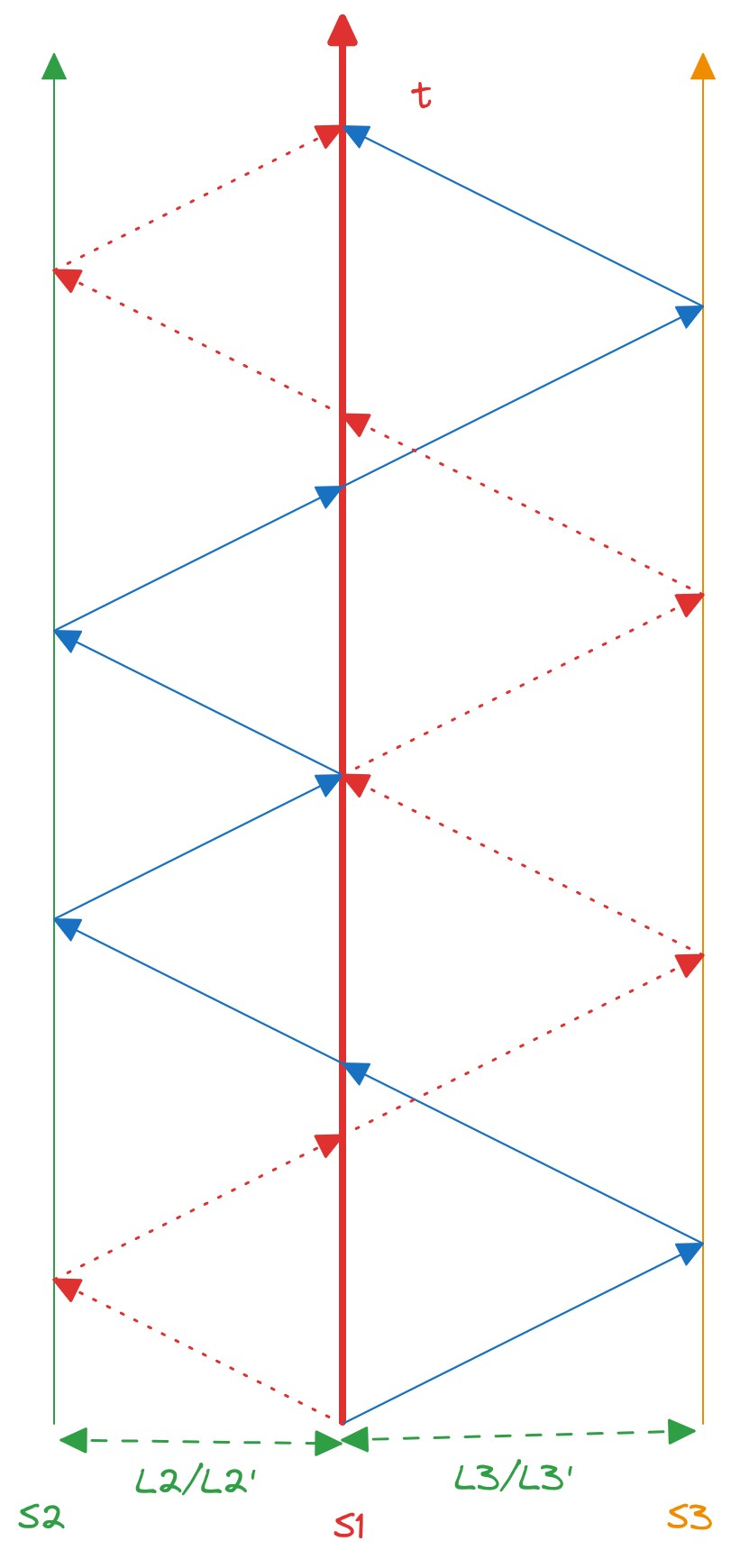}
    \caption{Space-time map of \ac{tdi} 2.0 for the Michelson-like X channel. }
    \label{fig:TDI-2.0}
\end{figure}

Suppose that all the armlengths are equal, i.e., $L_i = L$.
Thus, in the time domain, the first generation of TDI Michelson-like $X$ channel will be
\begin{equation}
    X_{1.0} =
    [y_{32'1} + \mcD y_{123}] + \mcD^2 [y_{231} + \mcD y_{13'2}]
    - [y_{231} + \mcD y_{13'2}] - \mcD^2 [y_{32'1} + \mcD y_{123}],
    \label{eq:TDI_Xt_yslr}
\end{equation}
where $\mcD = \mcD_i$ and $\mcD^2 = \mcD \mcD$.
Simply, let $y_{slr,nL} = y_{sr}(t -n L)$, its Fourier transform will be $\tdy_{slr,nL} = \mc{D}^n \tdy_{sr}$,
where $\mc{D}$ is the time delay.
Otherwise, one can easily get the Frequency domain TDI channel as
\begin{eqnarray}
    \tdX_{1.0} &=&  [\tdy_{31} + \mc{D} \tdy_{13}] + \mc{D}^2 [\tdy_{21} + \mc{D}\tdy_{12}] 
    - [\tdy_{21} + \mc{D}\tdy_{12}] - \mc{D}^2 [\tdy_{31} + \mc{D}\tdy_{13}] \nonumber \\
    &= & (1-\mc{D}^2) \left[ \tdy_{31} + \mc{D} \tdy_{13} - \tdy_{21} -\mc{D} \tdy_{12} \right],
    \label{eq:TDI_Xf_yslr}
\end{eqnarray}
Similarly, the second generation of TDI Michelson-like $X$ channel in the Frequency domain will be
\begin{eqnarray}
    \tdX_{2.0} &=& \, \tdy_{31} + \mcD \tdy_{13}
        + \mcD^2 \tdy_{21} + \mcD^3 \tdy_{12}
        + \mcD^4 \tdy_{21} + \mcD^5 \tdy_{12}
        + \mcD^6 \tdy_{31} + \mcD^7 \tdy_{13}
        \nonumber \\
        && - \tdy_{21} - \mcD \tdy_{12}
        - \mcD^2 \tdy_{31} - \mcD^3 \tdy_{13}        
        - \mcD^4 \tdy_{31} - \mcD^5 \tdy_{13}
        - \mcD^6 \tdy_{21} - \mcD^7 \tdy_{12}
        \nonumber \\
        &= & (1-\mcD^4) \tdX_{1.0}.
    \label{eq:TDI_2.0_fd}
\end{eqnarray}

However, different channels will use the same link, then the instrumental noises in different channels may be correlated with each other.
Considering that all the satellites are identical, we can get one ``optimal" combination by linear combinations of $X$, $Y$, and $Z$ \cite{Prince:2002hp}:
\begin{eqnarray}
    A & = \frac{1}{\sqrt{2}} (Z -X),
    \label{eq:A_XYZ} \\
    E & = \frac{1}{\sqrt{6}} (X -2Y +Z),
    \label{eq:E_XYZ} \\
    T & = \frac{1}{\sqrt{3}} (X + Y +Z).
    \label{eq:T_XYZ}
\end{eqnarray}
In the $A$, $E$, and $T$ channels, the instrumental noise is orthogonal,
and consequently, the noise correlation matrix of these three combinations is diagonal~\cite{Prince:2002hp}.
\subsection{Instrument noise}


We will focus on the case where the instrumental noise $n(t)$ is assumed to be Gaussian stationary with a zero mean.
Thus, the ensemble average of the Fourier components of the noise $n(f)$ can be written in the following form
\begin{equation}
    \left\langle \tilde{n}(f) \tilde{n}^*(f') \right\rangle = \frac{1}{2} \delta(f-f')S_n(f),
    \label{eq:f_tnoise}
\end{equation}
where ${}^*$ denotes complex conjugate, and $S_n(f)$ is the single-sided noise \ac{psd}\footnote{Because $n(t)$ is real, $\tilde{n}^*(f) = \tilde{n}(-f)$ and therefore $S_n(-f) = S_n(f)$.}.

For TianQin, the designed requirement for the acceleration noise is $\sqrt{S_a} = 10^{-15} {\rm m \, s}^{-2} {\rm Hz}^{-1/2}$ and the displacement noise is $\sqrt{S_x} = 1 {\rm pm \, Hz}^{-1/2}$ \cite{Luo:2015ght}.
For \ac{lisa}, as reported in Ref.~\cite{Babak:2021mhe}, the displacement noise is $\sqrt{S_x} = 15 \,{\rm pm}\, {\rm Hz}^{-1/2}$ and for the acceleration noise is $\sqrt{S_a} = 3\times 10^{-15} {\rm m}\, {\rm s}^{-2}{\rm Hz}^{-1/2}$.
For Taiji, the design goal for the displacement noise is $\sqrt{S_x} = 8 \,{\rm pm}\, {\rm  Hz}^{-1/2}$ and for the acceleration noise is $\sqrt{S_a} = 3 \times 10^{-15} {\rm m}\, {\rm s}^{-2}{\rm Hz}^{-1/2}$ at 1 \ac{mhz} \cite{Ruan:2020smc}.

As discussed at the beginning of section~\ref{sec:TDI}, when the laser noise is canceled, the total can be described by two noises.
One is displacement or position noise, which is dominated at high frequencies.
The other one is the acceleration noise, which is dominated at low frequencies.
Note that the noise parameters defined in the previous paragraph should convert to the same dimension, such as in the dimension of length (here, using the \ac{lisa} noise as an example)
\begin{eqnarray}
    \sqrt{S^{oms}_{\delta l}} (f) =& \sqrt{S_x} \sqrt{1 + \left( \frac{2 \rm mHz}{f} \right)^4}, 
    \\
    \sqrt{S^{acc}_{\delta l}} (f) =& \frac{\sqrt{S_a}}{(2\pi f)^2} \sqrt{1+\left( \frac{0.4 \rm mHz}{f} \right)^2} \sqrt{1+\left(\frac{f}{8 \rm mHz}\right)^4}.
\end{eqnarray}
and in the dimension of the relative frequency, it will be
\begin{eqnarray}
    \sqrt{S^{oms}_{\delta \nu / \nu}}(f) = & \sqrt{S_x}  \frac{2\pi f}{c} \sqrt{1 + \left( \frac{2 \rm mHz}{f} \right)^4}, 
    \\
    \sqrt{S^{acc}_{\delta \nu/\nu}} (f) =& \frac{\sqrt{S_a}}{2\pi f c}
    \sqrt{1+\left( \frac{0.4 \rm mHz}{f} \right)^2} \sqrt{1+\left(\frac{f}{8 \rm mHz}\right)^4} .
\end{eqnarray}
For different detectors, the difference is the value at the front and the tail of the frequency variation.
For TianQin, the relative noise parameters will be \cite{Luo:2015ght}
\begin{eqnarray}
    & \sqrt{S^{oms}_{\delta l}} (f) = \sqrt{S_x},
    \qquad
    \sqrt{S^{acc}_{\delta l}} (f) = \frac{\sqrt{S_a}}{(2\pi f)^2} \sqrt{1+ \frac{0.1 \rm mHz}{f} },
    \\
    & \sqrt{S^{oms}_{\delta \nu / \nu}}(f) =  \sqrt{S_x} \frac{2\pi f}{c},
    \quad
    \sqrt{S^{acc}_{\delta \nu/\nu}} (f) = \frac{\sqrt{S_a}}{2\pi f c}
    \sqrt{1+\frac{0.1 \rm mHz}{f}}.
\end{eqnarray}
With the above definitions and the assumption that all the instrument's noise parameters are the same, the PSD and cross-spectrum of noise for the TDI-1.0 in the $X, Y, Z$ channels will be \cite{armstrong_time-delay_1999, Babak:2021mhe}
\begin{eqnarray}
    S_n^{X,1.0} =& 16 \sin^2(f/f_*) \left[ S^{oms} + 2(1+\cos^2(f/f_*)) S^{acc} \right],
    \\
    S_n^{XY,1.0} =& -8 \sin^2(f/f_*) \cos(f/f_*) \left( S^{oms} + 4 S^{acc} \right),
\end{eqnarray}
where $f_* = c/(2\pi L)$.
For the $A, E, T$ channels, the PSDs will be \cite{armstrong_time-delay_1999, Babak:2021mhe}
\begin{eqnarray}
    S_n^{A, E}(f) &= & 8 \sin^2\left(f/f_*\right)
        \big\{ 4[1+\cos{(f/f_*)} + \cos^2{(f/f_*)}] S^{acc}
        \nonumber \\
        && + [2+\cos{(f/f_*)}]S^{oms} \big\},\\
        S_n^T(f) &= &
        32\sin^2{(f/f_*)}\sin^2{(f/f_*/2)}
        \left[4\sin^2{(f/f_*/2)} S^{acc}+S^{oms}\right].
    \label{eq:AE_snr}
\end{eqnarray}
And the noise PSD of the second-generation TDI channels is
\begin{eqnarray}
    S_n^{\mathcal{I},2.0} =& 4\sin^2(2 f/f_*) S_n^{\mathcal{I},1.0}
\end{eqnarray}
where $\mathcal{I} = X,Y,Z,XY, \ldots, A,E,T$.

\begin{figure}
    \centering
    \includegraphics[width=0.6\linewidth]{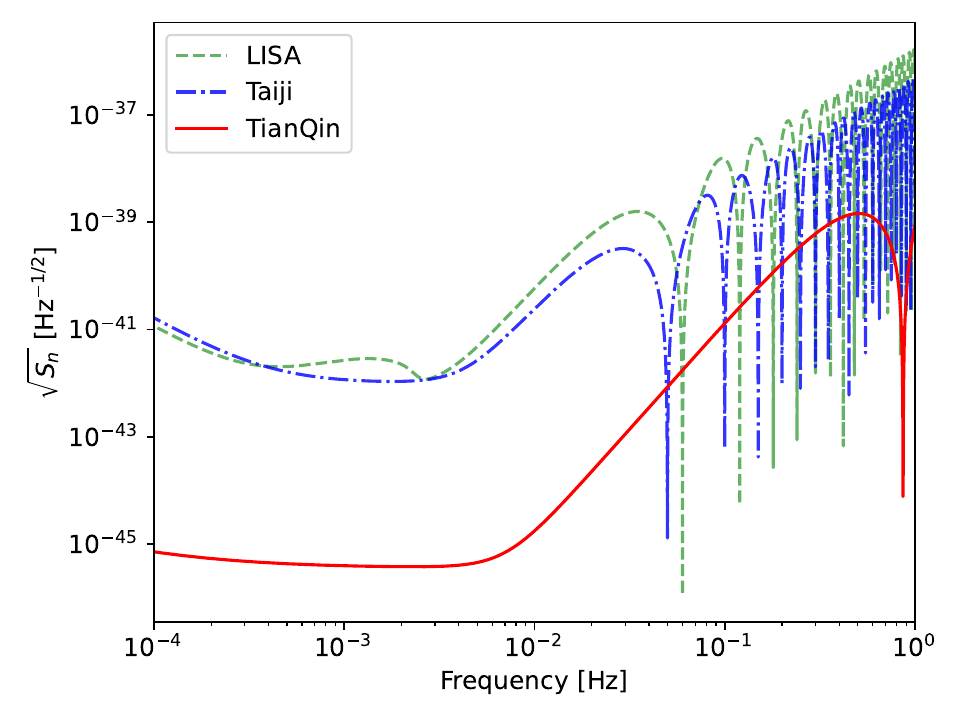}
    \caption{Noise \ac{psd} of TDI $A$ channel for \ac{lisa}, Taiji, and TianQin (four year data).}
    \label{fig:TDI_A_PSD}
\end{figure}

In Fig.~\ref{fig:TDI_A_PSD}, we have shown the three noise \ac{psd} curves of TDI first generation $A$ channel for \ac{lisa}, Taiji, and TianQin.

\section{Waveform}
\label{sec:source}

To extract information from the detector data, one should model the entire detection process.
With the basic definition in section \ref{sec:basic_def}, one can build a model for some general \ac{gw} signals.
To know the type of \ac{gw} source and more information about the \ac{gw} systems, an exact waveform is needed.
In this section, we review some waveforms we use for each type of source.

\subsection{Galaxy Compact Binary}
\label{sec:GCB}

In the \ac{mhz} frequency band, \ac{gw} events are mainly composed of \ac{wdb} in the Milky Way (with the number $\sim \mathcal{O}(10^8)$) \cite{Nelemans:2001hp}, which are expected to be the most numerous \ac{gw} sources for \ac{sbd}.
These \acp{gcb} are expected to exhibit relatively little frequency evolution.
Thus, the GW strain emitted from a \ac{gcb} can be safely approximated as (in the source frame) \cite{Katz:2022yqe} 
\begin{eqnarray}
    h_+ (t) &=& A_+ \cos\Phi(t)
    = h_0 \frac{1+\cos^2\iota}{2} \cos\Phi(t),
    \label{eq:hp}\\
    h_\times(t) &=& A_\times \sin\Phi(t)
    = h_0 \cos\iota \sin\Phi(t),
    \label{eq:hc}\\
    h_0 &=& \frac{4 (G \mathcal{M}_c)^{5/3}}{c^4 D_L} (\pi f)^{2/3},
    \label{eq:h0}\\
    \Phi(t) &=& 2 \pi f t + \pi \dot{f} t^{2} + \frac{\pi}{3} \ddot{f} t^2 + \phi_{0},
    \label{eq:Phi}
\end{eqnarray}
where 
$\iota$ is the inclination angle of the quadruple rotation axis with respect to the line of sight (here the direction is from the source to the Sun),
$\mathcal{M}_c = (m_1 m_2)^{3/5} / (m_1 + m_2)^{1/5}$ is the chirp mass of the system
($m_1$ and $m_2$ are the individual masses of the components of the binary),
$D_L$ is the luminosity distance to the source,
$\phi_0$ is the initial phase at the start of the observation,
$f$, $\dot{f}$, and $\ddot{f}$ are the frequency of the source,the frequency's derivative, and the double derivative with respect to time,
and $\ddot{f} = \frac{11}{3} \frac{\dot{f}^2}{f^2}$.

Considering the motion of the detectors moving around the Sun, a Doppler modulation of the phase of the waveform should be taken into account, i.e.,
\begin{eqnarray}
    \Phi(t) &\rightarrow \Phi(t) + \Phi_D(t), \\
    \Phi_D(t) &= 2\pi (f +\dot{f} t)\, \frac{R}{c} \cos\beta\cos(2\pi f_m t - \lambda),
    \label{eq:gcb_doppler}
\end{eqnarray}
where $\Phi_D(t)$ is the Doppler modulation, 
$f_m=1/$year is the modulation frequency,
$\beta$ and $\lambda$ are the latitudes and the longitude of the source in ecliptic coordinates,
$R$=1AU is the semi-major axis of the guiding center of the satellite constellation, respectively.
However, one should note that the Doppler modulation shown in Eq.~(\ref{eq:gcb_doppler}) needn't be considered if we use the response defined in the \ac{ssb} coordinate and the response function defined in Sec.~\ref{sec:midly_chirp}.
Because $\bfk \cdot \bfp_0$ will be the Doppler effect.

\subsection{Black Hole Binary}
\label{sec:BHB}

\paragraph{General Phenomenological waveform}

For a \ac{bhb} system, one can describe its waveform in the time domain or frequency domain with the help of the stationary phase approximation.
Here, we consider the frequency domain \texttt{IMRPhenomD} waveform, which assumes aligned spin, so only two parameters are needed to describe the spin parameters \cite{Khan:2015jqa, Husa:2015iqa}.
In this frame, a \ac{bhb} system can be characterized by four intrinsic parameters: masses $(m_1, m_2)$ and dimensionless spins $(\chi_1,\chi_2)$;
seven extrinsic parameters: luminosity distance $D_L$, inclination angle $\iota$, polarization angle $\psi$,
coalescence time and phase $(t_c, \phi_c)$
and the ecliptic longitude and ecliptic latitude $(\lambda,\beta)$ in the \ac{ssb}.
In the \texttt{IMRPhenomD} waveform model, the waveforms of plus and cross modes will be
\begin{eqnarray}
    \tdh_+(f) &=&
    \frac{\mathcal{M}_c^{5/6}}{\pi^{2/3}D_L}
    \frac{1 + \cos^2\iota}{2} f^{-7/6} \exp(\rmi\Psi(f)),
    \\
    \tdh_{\times}(f) &=&
    - \rmi \frac{M_c^{5/6}}{\pi^{2/3}D_L}
    \cos{\iota} \, f^{-7/6} \exp(\rmi\Psi(f)).
    \label{equ:imr waveform}
\end{eqnarray}
More details about the phase $\Psi(f)$ can be seen in Ref. \cite{Khan:2015jqa}.

\paragraph{Eccentric waveform}

The \ac{gw} emission causes the circularization effect, which makes the binaries almost non-eccentric when they are in the \ac{gbd} frequency band.
But when the binaries are in the \ac{sbd} frequency band, the eccentricity should be taken into account. Many eccentric waveform models have been developed to date \cite{Loutrel2020}. 
Here we use \texttt{EccentricFD}, which is a frequency-domain third post-Newtonian (3PN) waveform with initial eccentricity $e_0$ valid up to 0.4 \cite{Yunes2009, Huerta2014}, and has been included in \texttt{LALSuite}\cite{LSC2018}. 
This analytic model only contains the inspiral process of a binary, however, it is sufficient for \acp{sbhb}, as they are likely to merge outside the sensitive frequency band of \acp{sbd}.

Note that the \ac{bhb} system can be divided into \ac{mbhb} and \ac{sbhb} systems according to their masses and origin.
The heavier \ac{bhb} systems have lower frequency bands.
Though their origin or characteristics are different, their waveform formulas are similar.
When analyzing the data, it is important to note the range of parameter values and the applicability of the waveform.

\subsection{Extreme Mass Ratio Inspirals}
\label{sec:EMRI}

A prominent source that space-based detectors will detect is extreme mass ratio inspirals (EMRIs). These sources are formed by a stellar-mass compact object such as a stellar origin black hole ( $\mathrm{SOBH})$ with a mass $\mu \sim 1-10^2 \mathrm{M}_{\odot}$ inspiralling into a massive black hole $(\mathrm{MBH})$ with a mass $M \sim 10^4-10^7 \mathrm{M}_{\odot}$\cite{Fan:2020zhy, Ye:arxiv2023}.
Specifically, the time domain dimensionless strain of an EMRI source $h(t)$ can be given by 
\begin{equation}
    h(t)=\frac{\mu}{d_L} \sum_{l m k n} A_{l m k n}(t) S_{l m k n}(t, \theta) e^{i m \phi} e^{-i \Phi_{m k n}(t)}, 
\end{equation}
where $t$ is the time of arrival of the gravitational wave at the solar system barycenter, 
$\theta$ is the source-frame polar viewing angle, 
$\phi$ is the source-frame azimuthal viewing angle, 
$d_L$ is the luminosity distance, 
and $\{l, m, k, n\}$ are the indices describing the frequency-domain harmonic mode decomposition\cite{2021PhRvD.104f4047K,Chua:2020stf}. 
The indices $l, m, k$, and $n$ label the orbital angular momentum, azimuthal, polar, and radial modes, respectively. 
$\Phi_{m k n}=m \Phi_{\varphi}+k \Phi_\theta+n \Phi_r$ is the summation of decomposed phases for each given mode. 
The amplitude $A_{l m k n}$ is related to the amplitude $Z_{l m k n}^{\infty}$ of the Teukolsky mode amplitude far from the source. 
It is given by $A_{l m k n}=-2 Z_{l m k n}^{\infty} / \omega_{m k n}^2$, where $\omega_{m k n}=m \Omega_{\varphi}+k \Omega_\theta+n \Omega_r$ is the frequency of the mode, and $\Omega_{r, \theta, \phi}$ describe the frequencies of a Kerr geodesic orbit.

To expedite the generation of EMRI signals, we utilize the FastEMRIWaveform (FEW) package\footnote{\url{https://github.com/BlackHolePerturbationToolkit/FastEMRIWaveforms}}. The FEW is optimized to generate gravitational wave signals efficiently with GPU acceleration \cite{Chua:2020stf,chua2019reduced}. 
The trajectory calculation is done through a large-scale cubic-spline interpolation of $A_{l m n}(t)$ and $\Phi_{r, \varphi}(t)$ at a sparse set of times ( $\sim 10^2$).
A reduced-order-model technique is employed in actuality, reducing the number of harmonic modes needed by approximately 40 times and thereby significantly cutting down the time needed to generate the waveform for each source \cite{chua2019reduced,2021PhRvD.104f4047K}.
For example,$l \in [2,10]$,$m \in [0,l]$ and $n \in [-30,30]$,which totals 3843 modes reduce to $\sim 10^{2}$ modes.
So far, the fully relativistic FEW model is limited to eccentric orbits in the Schwarzschild spacetime.

\subsection{Stochastic Gravitational Waves Background}
\label{sec:SGWB}

In addition to the aforementioned primary distinguishable \ac{gw} sources, there is another important type of \ac{gw} source that could potentially be detected by \ac{sbd}, known as the \ac{sgwb}.
\ac{sgwb} is composed by a huge number of independent and unresolved \ac{gw} sources \cite{Romano:2016dpx}.
These stochastic signals are effectively another source of noise in \ac{gw} detectors.
A \ac{sgwb} can be written as a superposition of plane waves with  frequencies of $f$ and coming from different directions $\htk$ on the sky
\begin{equation}
    h_{ij}(t,\bfx) = \sum_P 
    \int_{-\infty}^{+\infty} df
    \int_{S^2} d\Omega_{\htk}
    \tdh_{P} (f,\htk) e^{P}_{ij}(\hat{k})
    e^{\rmi 2\pi f [t-\htk \cdot \bfx (t)/c]},
\end{equation}
where $P = \{+,\times \} $ denotes polarization.
As a stochastic source, one can treat the complex amplitude $\tdh_P(f, \htk)$ as some random variable with zero mean value.
Supposing the \ac{sgwb} is stationary, Gaussian, isotropic, and unpolarized, the ensemble average of the two random amplitudes $\tdh_P(f,k)$ can be defined as \cite{Romano:2016dpx, Allen:1997ad}
\begin{equation}
    \langle \tdh_{P}(f,\hat{k})\tdh^{*}_{P'}(f',\hat{k}')\rangle
    = \delta(f-f')
    \frac{\delta^2 (\htk, \htk')}{4\pi}
    \delta_{PP'}
    \frac{1}{2} S_h(f).
\end{equation}
The function $S_h(f)$ is the one-sided \ac{psd} of \ac{sgwb}.

Note here that $\delta^2(\htk, \htk')$ is a Dirac delta over the two-sphere, and it implies that the \ac{sgwb} is independent of $\htk$.
However, it is expected that the \acp{sbd} will detect millions of \acp{wdb} in the Milky Way and nearby universe \cite{LISA:2022yao}, and the superposition of millions of unresolved \acp{wdb} will contribute to an \ac{sgwb} \cite{Rieck:2023pej} (often referred to as foreground due to its strength).
Furthermore, due to our location at one end of the Milky Way, this \ac{sgwb} is anisotropic.
Of course, there may exist other anisotropic \acp{sgwb} as well \cite{LISACosmologyWorkingGroup:2022kbp}.
In this case, the \ac{psd} of the anisotropic \ac{sgwb} will depend on the frequency and direction as $\mathcal{P}(f, \htk)$.
If we assume \ac{sgwb} is directional and frequency independent, the \ac{psd} can be factorized as \cite{Allen:1996gp}
\begin{equation}
    \mathcal{P} (f,\hat{k}) = H(f)\mathcal{P}_h(\hat{k})
\end{equation}
where the \ac{psd} of the \ac{sgwb} is given by $H(f)$, and the $\mathcal{P}_h(\hat{k})$ describes the distribution of signal.




\section{Example data-set}
\label{sec:data_set}

To simulate the joint observation of certain \ac{gw} signals, it is necessary to have precise knowledge of the relative positions of the three detectors.
The relative positions of guiding centers for each detector can be determined by the initial phase parameter $\alpha -\beta$ or $\kappa_0$ and $\alpha''$ as defined in Eqs.~(\ref{eq:earth_x})-(\ref{eq:earth_z}) and Eqs.~(\ref{eq:lisa_x})-(\ref{eq:lisa_z}).
Additionally, the relative position of the spacecrafts in different detectors can be determined by the initial phase of the spacecraft (here is the initial phase parameter $\lambda$ and $\lambda'$).
Once the detectors are launched, the relative phase and positions are fixed.
However, when simulating data for testing purposes, the initial phase parameters are some arbitrary values.

\begin{figure}[htbp]
    \centering
    \includegraphics[width=0.48\linewidth]{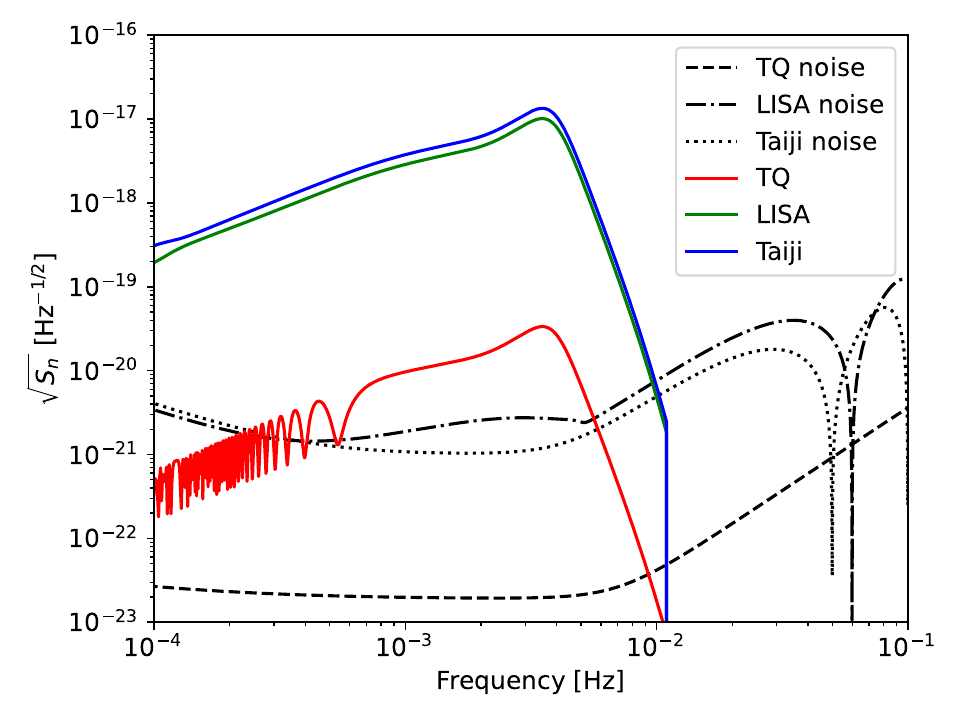}
    \includegraphics[width=0.48\linewidth]{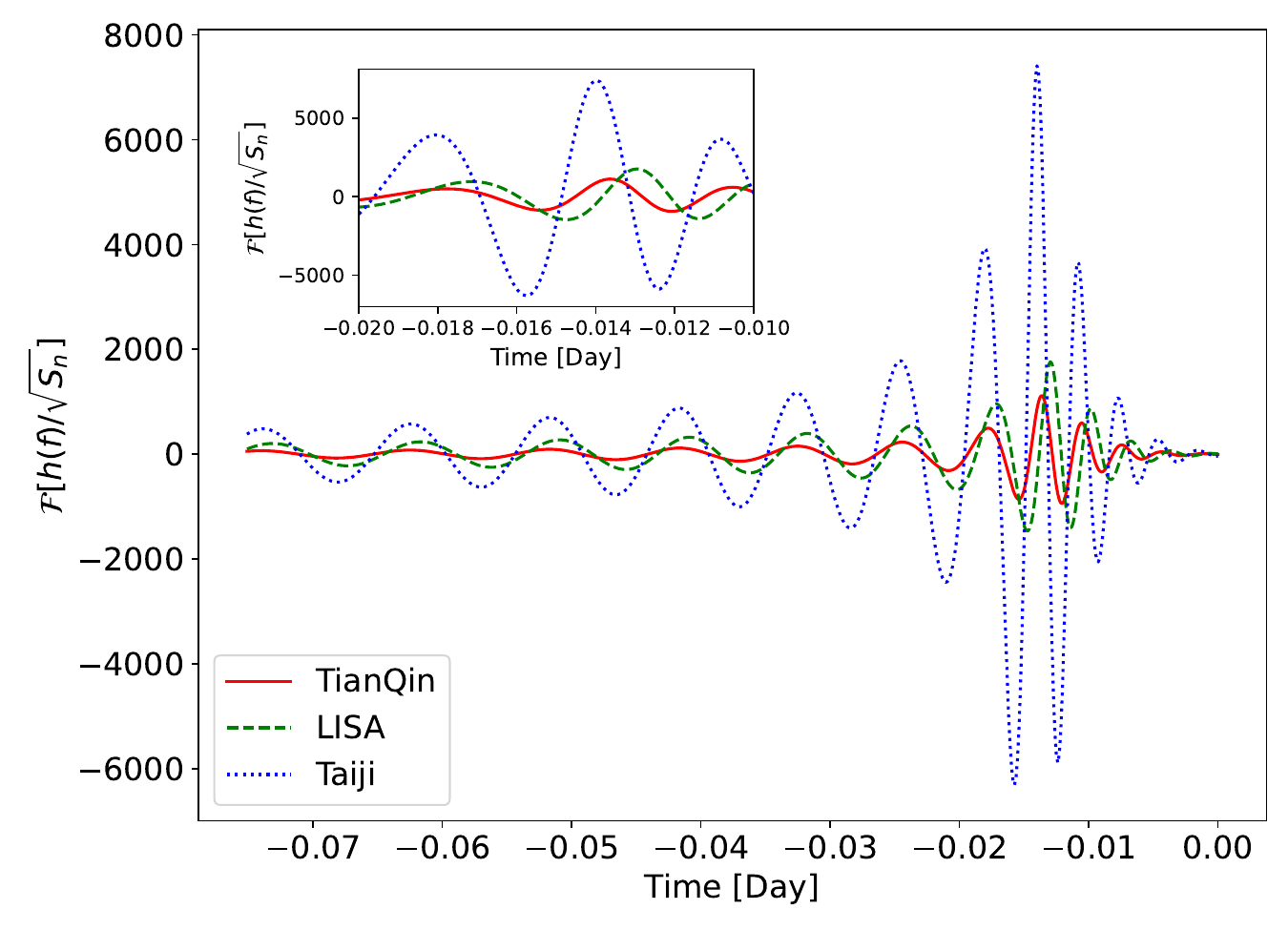}
    \caption{Left: TDI-A channel responsed \ac{mbhb} signal and relative noise \ac{psd} detected by different detectors in frequency domain.
    Right: The signals in time domain, and the signal here is the strain divided by the noise ASD, which can make the responsed signals within similar magnitude.
    The masses for the binary system are ($3.5\times 10^6, 2.1\times 10^5$) $M_\odot$, spins are $(0.2, 0.1)$, the luminosity distance is $10^3$ Mpc, the position is $(\lambda, \beta) = (0.4,1.2)$, and $\iota = 0.3, t_c =
    0$.
    The total observation time is three months.
    Here, the \texttt{IMRPhenomD} waveform is applied.
    The initial phase of TianQin's and \ac{lisa}'s first spacecraft is set to 0 for this figure.
    }
    \label{fig:MBHB_fd}
\end{figure}

\ac{mbhb} are the primary sources for \acp{sbd}, and its total inspiral-merger-ringdown phase can be detected in the \ac{mhz} band.
In Fig.~\ref{fig:MBHB_fd}, we have shown the \ac{mbhb} event detected by TianQin, \ac{lisa}, and Taiji and relative noise \ac{psd}.
From the left figure, it can be found that the length of the arms gives \ac{lisa} and Taiji an advantage in terms of the response intensity to signals, but at the same time, it also results in higher low-frequency noise levels.
From the right panel, one can find that the amplitude and the phase of the response signal of the three detectors varies.
From the different phases, one can say that the signal will arrival at different detectors at different time.
This is the key feature of the joint simulating of \ac{gw} signals.
The time delay between the arrival of the signals from different detectors can be calculated by the relative positions of the detectors and the relative positions of the source.

\begin{figure}[htbp]
    \centering
    \includegraphics[width=0.9\linewidth]{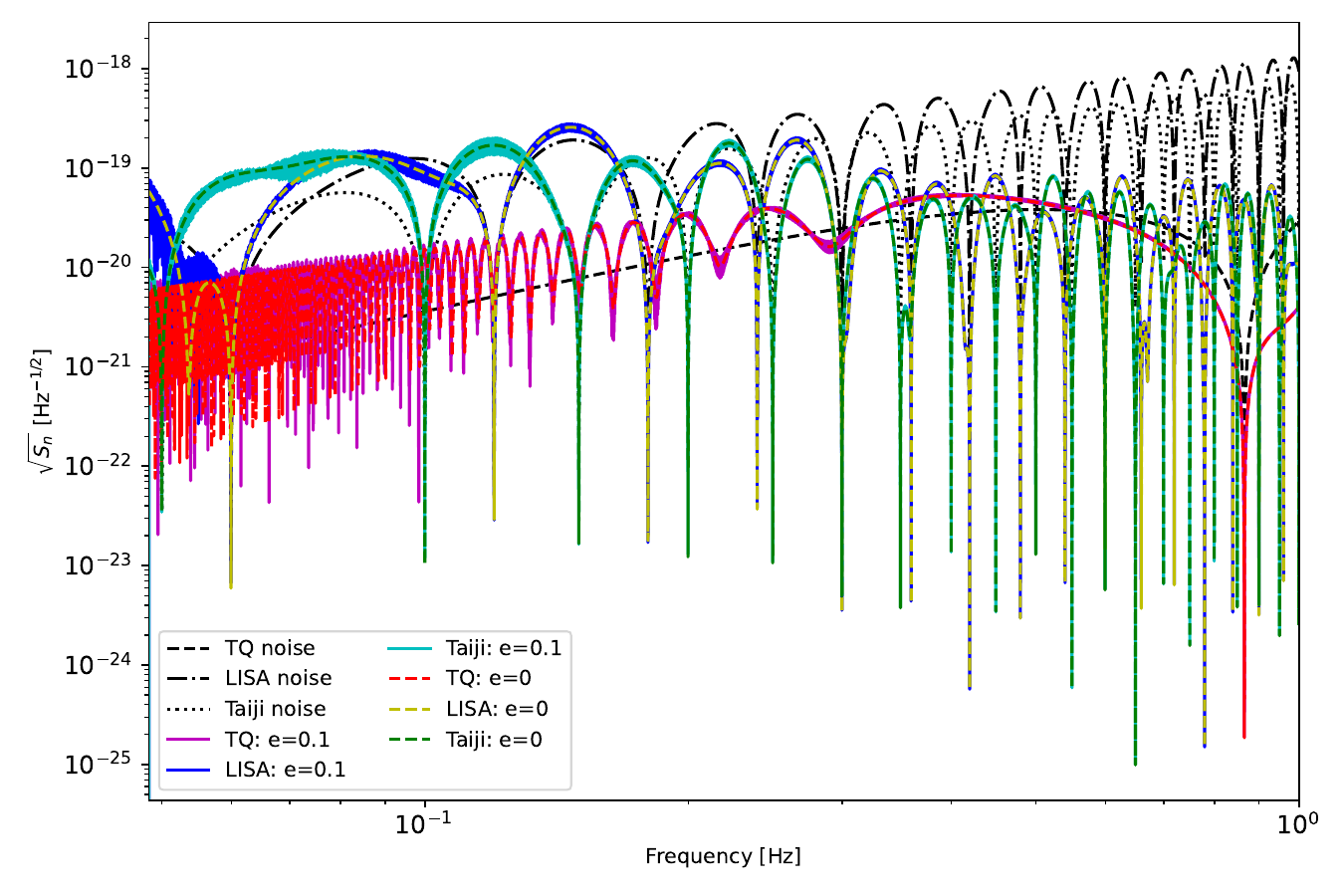}
    \caption{TDI-A channel responsed \ac{sbhb} signal and relative noise \ac{psd} detected by different detectors.
    The masses for the binary system are ($35.6, 30.6$) $M_\odot$, the luminosity distance is $100$ Mpc,
    the position is $(\lambda, \beta) = (4.7,-1.5)$,
    and $\iota = 0.3, t_c = 0$.
    The total observation time is three months.
    Here we have used the \texttt{EccentricFD} waveform.
    The initial phase of TianQin's and \ac{lisa}'s first spacecraft is set to 0 for this figure.}
    \label{fig:SBHB_fd}
\end{figure}

The mass of \ac{sbhb} systems is relatively lighter compared to \ac{mbhb}, which leads to these systems predominantly producing signals in higher frequency ranges.
In the Fig.~\ref{fig:SBHB_fd}, we can observe the performance of a \ac{sbhb} signal across different detectors.
Interestingly, when eccentricity is taken into account, the response waveform becomes considerably more intricate compared to the case where eccentricity is disregarded.
This increased complexity in the waveform poses significant challenges for data processing and analysis.
Furthermore, the figure demonstrates that the intersection point between the curve of the response signal from TianQin and the noise \ac{psd} is noticeably higher than in the case of \ac{lisa} or Taiji.
This observation suggests that TianQin exhibits certain advantages in high-frequency detection.

In the low-frequency region of Fig.~\ref{fig:MBHB_fd}, the post-response signal of TianQin shows oscillations.
Likewise, in Fig~\ref{fig:SBHB_fd}, the response signals from all three detectors demonstrate oscillatory behaviour.
These oscillations arise as a result of the orbital motion of the detectors.

\section{Summary}
\label{sec:sum}

Around 2035, one may see more than one \acp{sbd} operating simultaneously, with potential candidates including TianQin, LISA, and Taiji. 
Apart from the huge prospect of the scientific return from the joint observation over single detectors \cite{Torres-Orjuela:2023hfd}, there are also challenges in doing data analysis for joint observation. 
To facilitate the study of problems involved in the joint data analysis, we have introduced \texttt{GWSpace} in this paper, which is a package that can simulate the joint detection data from three \acp{sbd}: TianQin, LISA, and Taiji.

\texttt{GWSpace} uses \ac{ssb} as the common coordinate system for all detectors. 
It can simulate data for \ac{gcb}, \ac{bhb}, \ac{emri}, \ac{sgwb}, and simple burst signals. 
It supports injecting time-domain waveform functions and obtaining observed data through time-domain responses.
For frequency-domain waveforms, it supports the frequency-domain responses of regular 22 mode, higher harmonic modes, and waveforms with eccentricity \ac{bhb}.
The first TDI combinations are now included in the time domain and the frequency domain.
It includes the time-domain and frequency-domain responses of the 1st generation TDI combinations and the corresponding TDI noise.
We have also given some example data sets generated with the package.
The package is open source and is free for downloading and using\footnote{\href{https://github.com/TianQinSYSU/GWSpace}{https://github.com/TianQinSYSU/GWSpace}}. 
To clearly define all the notations and to eliminate possible misunderstanding, we have presented a detailed description of the coordinate system, the detector orbits, the detector responses, the \ac{tdi} combinations, the instrumental noise models, and the waveforms for each source in this paper.

As the first work in this direction, \texttt{GWSpace} can be further improved in many ways. For example, we have only implemented the first generation \ac{tdi} so far, while second-generation combinations are usually required, at least for LISA and Taiji. Moreover, a more robust response is needed for some sources with complex waveforms, such as \ac{bhb} systems with eccentricity \cite{Wang2023}.
The package still relies on very idealistic assumptions about the noise: the noises from all satellites in a detector are identical, while in reality, no two spacecraft can be the same \cite{Cheng:2022vct}. 

One can improve on the last point by implementing more sophisticated noise models for each detector, but the most precise noise model will have to come from the people responsible for each detector. We are hopeful that this may happen one day, and then \texttt{GWSpace} can serve as the starting point for a serious multi-mission data challenge for space-based \ac{gw} detection.

\section*{Acknowledgments}

This work has been supported in part by 
the National Key Research and Development Program of China 
(No. 2023YFC2206700),
the Guangdong Major Project of Basic and Applied Basic Research 
(Grant No. 2019B030302001), 
and the Natural Science Foundation of China 
(Grants No. 12173104 and No. 12261131504).
H. Fan is supported by the Hebei Natural Science Foundation with Grant No. A2023201041 and the Postdoctoral Fellowship Program of CPSF under Grant Number GZC20240366.
C.-Q. Ye is supported by the Sichuan University of Science \& Engineering Program (No. 2024RC031).
Several figures were created using \texttt{excalidraw}\footnote{\url{https://excalidraw.com/}}.

\section*{References}

\bibliographystyle{iopart-num}
\bibliography{GWSpace}
\end{document}